\documentclass[%
    reprint,
    showpacs,preprintnumbers,
    showkeys,
    amsmath,amssymb,
    aps,
    prx,
    floatfix,
    longbibliography,
    notitlepage,
]{revtex4-2}

\pdfoutput=1

\usepackage[pdftex]{graphicx}
\usepackage{tikz}
\usepackage{hyperref}
\hypersetup{
	colorlinks=true,
	urlcolor=blue,
	citecolor=blue,
	linkcolor=blue,
	breaklinks
}

\usepackage{mathtools}
\usepackage{wasysym}

\usepackage[utf8]{inputenc}
\usepackage[T1]{fontenc}
\usepackage{braket}
\usepackage[normalem]{ulem}
\usepackage{dsfont}
\usepackage{xcolor}
\usepackage{nicefrac}
\usepackage{enumerate}
\usepackage{float}
\usepackage{comment}
\usepackage{wrapfig}
\usepackage[caption=false]{subfig}

\definecolor{mygray}{gray}{0.5}

\begin{document}

\title{Spin Qubit Leapfrogging: Dynamics of shuttling electrons on top of another}

\author{Nicklas Meineke}
\email{nicklas.meineke@uni-konstanz.de}
\affiliation{Department of Physics and IQST, University of Konstanz, 78457 Konstanz, Germany}
\author{Guido Burkard}
\email{guido.burkard@uni-konstanz.de}
\affiliation{Department of Physics and IQST, University of Konstanz, 78457 Konstanz, Germany}

\begin{abstract}
    Spin shuttling has crystalized as a powerful and promising tool for establishing intermediate-range connectivity in semiconductor spin-qubit devices. Although experimental demonstrations have performed exceptionally well on different materials platforms, the question of how to handle areas of low valley splitting in silicon during shuttling remains unresolved. In this work, we explore the possibility of utilizing the valley degree of freedom, particularly in regions of low valley splitting, to allow mobile spin qubits to be shuttled through an occupied stationary quantum dot, thereby leapfrogging over the stationary electron. This not only grants a more enriched mobility for shuttled electrons, as it opens new possible routing paths, but also enables the implementation of an entangling SWAP$^\gamma$ two-qubit gate operation in the process. Simulating this process for different sets of parameters, we demonstrate the feasibility of such an operation and offer a unique use case for otherwise precarious regions of a quantum processor chip and propose a possible extension to the set of possible operations for silicon spin qubit devices.
\end{abstract}

\maketitle

\section{Introduction}
\label{sec:Introduction}
Semiconductor spin qubits offer an up-and-coming platform for implementing potentially large-scale quantum computers due to their long coherence times \cite{Coherence}, high fidelity operations \cite{highfid1,highfid2,highfid3} and their compatibility with  existing semiconductor fabrication technology \cite{fab1,fab2}. To realize large-scale qubit architectures, one requires methods for fast high-fidelity transport of quantum information through the chip \cite{coupler,architecture1}. While this can be achieved efficiently and with high fidelity through a number of SWAP-gates \cite{hrl2026}, spin shuttling has recently distinguished itself as a scalable method for interactions between distant qubits in semiconductor devices with fidelities above the error correction threshold \cite{highfid2,conveyor1,conveyor2,conveyor3,bucketbrigade1,bucketbrigade2}. Currently, the two most well-studied flavors of spin shuttling in silicon devices are the bucket-brigade shuttling, where the qubit is adiabatically transferred through an array of empty quantum dots \cite{bucketbrigade1,bucketbrigade2} and conveyor-belt shuttling, consisting of moving the electron by trapping it in a traveling-wave potential \cite{conveyor1,conveyor2,conveyor3}. Especially the conveyor belt method has received significant attention in recent times by presenting high shuttling fidelities over long distances $\sim 20 \,\mu {\rm m}$ \cite{20mumshuttle} and allowing the implementation of two-qubit gates between shuttled electron-spin qubits \cite{mobilegates}, while simultaneously minimizing the amount of necessary independent control lines. However, for conveyor belt shuttling, one particularly formidable challenge is the local and rapid variation of material parameters such as the valley splitting in silicon, creating the risk of detrimental spontaneous excitations into non-computational states and occurrence of decoherence \cite{conveyor1}. Although there have been advances in producing silicon wafers with a higher average valley splitting \cite{valleyengineering} and in mapping the valley splitting of a device \cite{map-valley1,map-valley2}, as well as proposals to circumnavigate precarious regions \cite{avoidvalley1,avoidvalley2,avoidvalley3}, it is still an open question how future architectures shall deal with this obstacle.  

While one usually seeks to avoid exciting a spin qubit into an excited valley state, in this work we present a method to utilize this additional degree of freedom to allow spin qubits to leapfrog over another by temporarily occupying excited valley states. We model the process of a mobile qubit encountering a stationary occupied quantum dot as a triple quantum dot (TQD) in which the middle dot is always filled by an electron in a ground valley state with arbitrary spin configuration. Thereby, this essentially describes a bucket-brigade shuttling through an already occupied dot, but can also be interpreted as describing a solitary fixed and occupied quantum dot placed in between two shuttling lanes in the conveyor belt mode. Through a detuning sequence, the mobile qubit is then loaded into the occupied dot, to be later unloaded into the empty quantum dot on the other side of the TQD. Because of the Pauli exclusion principle, the spin triplet components of the two-electron wave function will have one electron transition into an excited valley state in order for the two qubits to be able to occupy the same site. Consequently, the triplets will carry an additional energy equal to the valley splitting energy $E_m$ of the middle dot, resulting in the collection of a relative phase and the implementation of an entangling SWAP$^\gamma$-gate, where $\gamma$ is an arbitrary exponent which is tunable by the waiting time in the doubly-occupied charge configuration. Compared to a SWAP$^\gamma$ operation mediated by tunnel barrier modulation, this approach presents two major advantages: First, it does not require any adressability of the tunnel barriers, thus simplifying control overhead and relaxing architectural constraints related to wiring, and second, it can be and is best utilized in areas of very low valley splitting, where usual implementations of exchange gates become problematic \cite{Weinstein2023-cn}. 

To assess the feasibility of such a protocol, we will construct an effective Hamiltonian for the system, investigate how different error sources and quasistatic noise affect the procedure, and simulate the leapfrogging dynamics using QuTiP \cite{qutip} for two sets of realistic device parameters. We find that for very low valley splittings ($\approx 2 \mu $eV) in the central dot, this operation displays sufficient resilience against noise, allowing a novel utilization for these treacherous regions. This results in expected fidelities of $1-\mathcal{F} \approx 4.5\times 10^{-3}$ for the SWAP$^\gamma$ gates, where further improvements through pulse optimization can be expected.

\section{Model}\label{sec:Model}
\begin{figure*}[t]
    \centering    
    \includegraphics[width=1\linewidth]{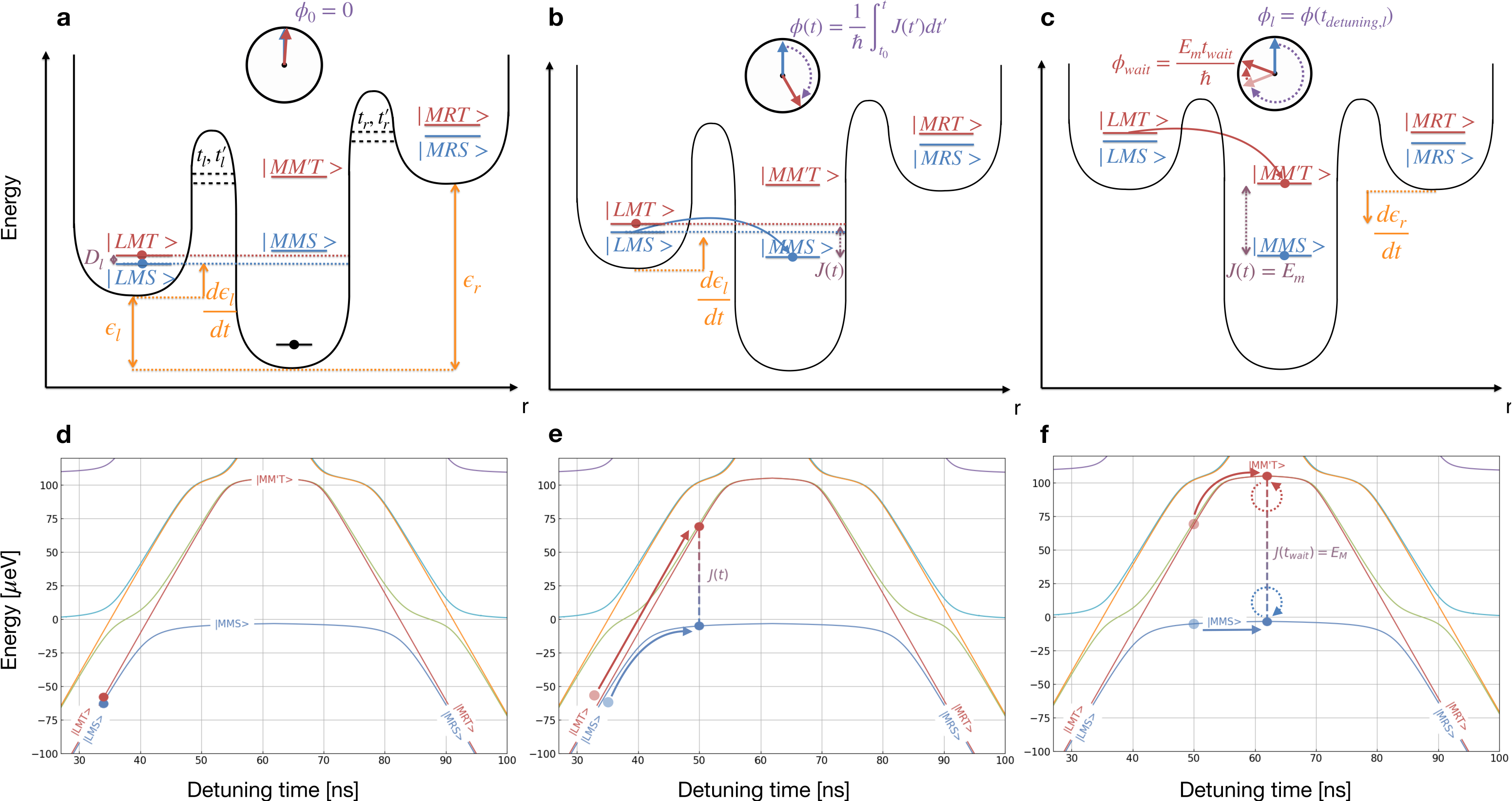}
    \caption{Schematic of the time evolution of the two qubits involved in the leapfrogging process. The top row (\textbf{a}-\textbf{c}) shows schematic potential energy diagrams as a function of position at three stages of the protocol, while the lower row (\textbf{d}-\textbf{f}) depicts the  instantaneous eigenenergies as a function of time with corresponding time and singlet (triplet) state indicated by a blue (red) dot. The setup is modelled through a TQD inhabited by two qubits, where one is bound to the middle dot due to its strong detuning and the other moves from the left to the right. For simplicity, we assume the spins are initialized in opposite directions whereby their full two-electron wave function is always a superposition of spin singlet and $T_0$ triplet state. (\textbf{a}, \textbf{c}) Initial state with mobile electron in the left dot. The detuning between the left and middle dots $\epsilon_l$ is still far enough from the charge transition that the $S-T_0$-splitting is negligible and the state collects no relevant relative phase. (\textbf{b}, \textbf{e}) The left dot is starting to be detuned. While the the singlet component transitions into the (0,2,0)-charge regime when it surpasses the charging energy $U-\frac{E_m}{2}$, the triplet component only transitions when the detuning matches the additional middle-dot valley splitting $E_m$. During this process, the triplet collects an additional detuning phase $\phi_l$ relative to the singlet. (\textbf{d}, \textbf{f}) After the charge transition is sufficiently complete, both spin configurations live in the (0,2,0)-charge regime where the $S-T_0$-splitting is constant $\Delta_{S-T_0} \approx E_m$. By waiting here, one can tune the relative phase of the triplet before preforming the inverse procedure on the right and collecting another detuning phase $\phi_r$.}
    \label{fig:TQD}
\end{figure*}
We model the process of a mobile qubit encountering a stationary occupied quantum dot as a triple quantum dot (TQD) in which the middle dot is always occupied by a qubit in an arbitrary spin state. The dots have valley splittings $E_l$, $E_m$, and $E_r$, going from left to right, the outer dots have a chemical potential  controlled by $\epsilon_l$ and $\epsilon_r$, and are coupled to the middle dot by tunneling parameters $(t_c)_l$, and $(t_c)_r$. We assume non-vanishing differences in valley phase $\theta_i$ of the outer dots $i \in \left\{ l,r\right\}$ in relation to the middle dot, whereby there exists a non-zero inter-valley coupling. Consequently, the coupling energy for transitioning from a valley ground state to a valley excited state during a charge transition and vice versa is given by $t'_{i} = \sin(\theta_i) (t_c)_{i}$, while the coupling energy for transitions preserving the valley state is given by $t_{i}  = \cos(\theta_i) (t_c)_{i}$.

At the start of the protocol, both electrons are expected to be in the valley ground state. It is assumed that the magnetic field gradients $\Delta B_\parallel$ and $\Delta B_\perp$ between the dots are vanishing and the spin-orbit coupling is negligible, whereby there is no coupling between different spin states (see Appendix~\ref{App:SO-Grad} for further discussion). The resulting full Hamiltonian of the system can therefore be written as follows,
\begin{equation}
    \begin{aligned}
        H &= \sum_{i}  \Bigg( U_1^{(i)} \hat{n}_i(\hat{n}_i-1) 
            + \sum_{j< i} U_2^{(i,j)} \hat{n}_i \hat{n}_j +  \epsilon_i \hat{n}_i \\
            &+\sum_{v \in \left\{+,-\right\}}  \frac{E_z^{(i)}}{2} \left(c_{i,\uparrow,v}^\dagger c_{i,\uparrow,v} - c_{i,\downarrow,v}^\dagger c_{i,\downarrow,v}\right) \Bigg)\\
            &+ \sum_{i,s} \lvert\Delta_i\rvert\left( e^{i\phi_i} c_{i,s,-}^\dagger c_{i,s,+} + e^{-i\phi_i}c_{i,s,+}^\dagger c_{i,s,-} \right) \\
            &+ \sum_{s,v} (t_c)_l \left( c_{l,s,v}^\dagger c_{m,s,v} + c_{m,s,v}^\dagger c_{l,s,v} \right) \\
            &+ \sum_{s,v} (t_c)_r \left( c_{m,s,v}^\dagger c_{r,s,v} + c_{r,s,v}^\dagger c_{m,s,v} \right) .
    \end{aligned}
\end{equation}
Here, $E_z^{(i)} = E_z$ is the Zeeman splitting of each respective dot, and $\hat{n}_i = \sum_{s \in \left\{ \uparrow, \downarrow \right\}}\sum_{v} c_{i,s,v}^\dagger c_{i,s,v} $ is the i-th occupation number operator. The on-site and nearest-neighbor charging energies $U_1^{(i)}=U_1$ and $U_2^{(i,j)}=U_2$ of the quantum dots are assumed to be equal everywhere, $\Delta_i $ denotes the complex-valued valley-coupling, and $E_i = \lvert \Delta_i\rvert $ and $\phi_i$ are its magnitudes and phases respectively.
Utilizing the Pymablock package \cite{pymablock}, we perform a Schrieffer-Wolff transformation to project out the (2,0,0), (0,0,2), and (1,0,1) charge states, which are typically unoccupied during the leapfrogging. Here, we refer to charge states of the TQD as ($n_L$,$n_M$,$n_R$) where $n_L$, $n_M$, and $n_R$ are the numbers of electrons in the left, middle, and right quantum dot. The resulting exchange couplings are on the scale of $\sim 10 \, {\rm neV}$, which is small enough to justify their omission. The diagonal energy corrections obtained in this way differ only slightly inside each respective charge sector and can thereby be absorbed into the detunings. The resulting Hamiltonian can then be split into four independent sectors, representing the uncoupled spin configurations $S$, $T_0$, and $T_\pm$. The Hamiltonians of the polarized triplets $T_\pm$ are equivalent to that of the unpolarized triplet $T_0$, but shifted by $\pm 2 E_z$. Consequently, we will focus our analysis onto the $S$ and $T_0$ Hamiltonians, as the polarized triplets follow analogously. The full Hamiltonians in matrix form and a more rigorous derivation can be found in Appendix~\ref{App: The Hamiltonian}. 

We assume that the mobile qubit arrives from the left of the stationary dot in an arbitrary spin state. Therefore, the initial state has the form $\ket{\psi(0)} = \alpha \ket{LMS} + \sum_s \beta_s \ket{LMT_s}$, where we use $\ket{LMS}$ to denote a state with one electron in the left ($L$) and the other in the middle ($M$) dot, where the two-electron spin state is a singlet ($S$), and analogously for the spin triplets $T_s$. We shift the definition of $\epsilon_l$, and $\epsilon_r$ such that $\epsilon_i = 0$ is satisfied at $U_1+\frac{E_m}{2}$, in order to have their zero value centered between the two relevant avoided crossings.

The protocol is initialized with the left dot being detuned negatively by $(\epsilon_l)_{\text{start}} \ll -|t_l|$ and the right dot being detuned positively by $(\epsilon_r)_{\text{start}} \gg |t_r|$, such that the initial (1,1,0)-charge configuration is  energetically strongly favorable. Then, while keeping $\epsilon_r$ constant,  $\epsilon_l = (\epsilon_l)_\text{start} + vt$ is increased linearly with a detuning rate $v$, until the desired detuning $(\epsilon_l)_\text{end}$ is reached. As can be seen in Figure~\ref{fig:TQD}, during this detuning process each of the (1,1,0)-states will encounter an avoided crossing with a (0,2,0)-state and adiabatically transition into these charge states under the Landau-Zener condition  $v \ll (t_c)_l^2$ (we use units where $\hbar=1$).
In this process, the singlet state $\ket{LMS}$ will transition to $\ket{MMS}$ comprising two electrons in the ground-state orbital of the middle QD, with the crossing occurring around $\epsilon_l =-\frac{E_m}{2}$ and the triplet states $\ket{LMT_s}$ will transition to $\ket{MM'T_s}$ with their crossing located at $\epsilon_l = \frac{E_m}{2}$. Therefore, the triplet states need to transition into a state with an additional valley excitation, since the Pauli exchange interaction forbids the existence of a $\ket{MMT_s}$-state. Consequentially, the singlet will have its energy reduced by $E_m$ in comparison to the triplet states, leading to an additional phase $(\phi_\text{detuning})_l$ collected by the triplets during the detuning process. In addition to a deterministic shift due to the Zeeman energy, this phase is the same for all triplets and can roughly be estimated as 
\begin{equation}
    \begin{aligned}
        &(\phi_\text{detuning})_l \approx \frac{1}{2v} \int_{(\epsilon_l)_\text{start}}^{(\epsilon_l)_\text{end}} \Bigg( E_m + \\
        &\sqrt{\left( \epsilon + \frac{E_m}{2} \right)^2 + 8t_l^2} - \sqrt{\left( \epsilon - \frac{E_m}{2} \right)^2 + 4t_l'^2}  \Bigg) d\epsilon.
    \end{aligned}    
\end{equation}
Before detuning the right dot to transport the mobile qubit to the other side, one can wait in this particularly charge-noise-protected configuration \cite{ErcanCoppersmith22} to collect an additional phase which is roughly equal to the wait time multiplied by the middle-dot valley splitting,
\begin{equation}   
    \phi_\text{wait} \approx E_m t_\text{wait}  ,  
\end{equation}
thereby tuning the phase which the triplet components of the quantum state collect during the protocol.

Repeating the inverse process for the right dot, the triplets collect another phase $(\phi_\text{detuning})_r$, whereby the final action on of the protocol on the initial states can be formulated as:
\begin{gather}
    \ket{LMS} \rightarrow -\ket{MRS}, \\
    \ket{LMT_s} \rightarrow - e^{-i\left((\phi_\text{detuning})_l + (\phi_\text{detuning})_r + \phi_\text{wait}\right)} \ket{MRT_s},
\end{gather}
where the minus signs originate from the fermionic exchange statistics and can be ignored as they present a global phase in the fully transitioned regime. It is important to note here that even though not explicitly included in the Hamiltonian, excited valley states have a slightly different g-factor $g + \delta g$ compared to their valley ground-state counterparts\cite{valleyphasemapping,map-valley2}. While this has no influence on the $(S_z)_{\text{tot}}=0$ configurations, the polarized triplets $\ket{T_-},\ket{T_+}$ will, as result, acquire additional deterministic detuning phases $\pm(\delta \phi_\text{detuning})_i$, as well as a wait-time dependent phase
\begin{equation}
    \varphi_\pm = \pm \mu_B B_zt_{\text{wait}}\delta g.
\end{equation}
As these phases are deterministic and equal but opposite in sign for the polarized triplets, they can easily be corrected for by a balanced $Z$-gate on both qubits, which can be implemented digitally. Assuming that this is done, the collected phases of all the triplets will coincide, leading to isotropic exchange.
As a result, the leapfrogging protocol, while transporting the mobile qubit from the left dot to the right dot, has the triplet components collect a total phase $\pi \gamma = (\phi_\text{detuning})_l + (\phi_\text{detuning})_r + \phi_\text{wait}$ and thereby implements a SWAP$^{\gamma}$ gate, which is tuned by waiting in the (0,2,0) charge configuration and can be used to implement arbitrary 2-qubit gates between the stationary and the mobile qubit.

\section{Results}
\begin{figure}
    \centering
    \includegraphics[width=1\linewidth]{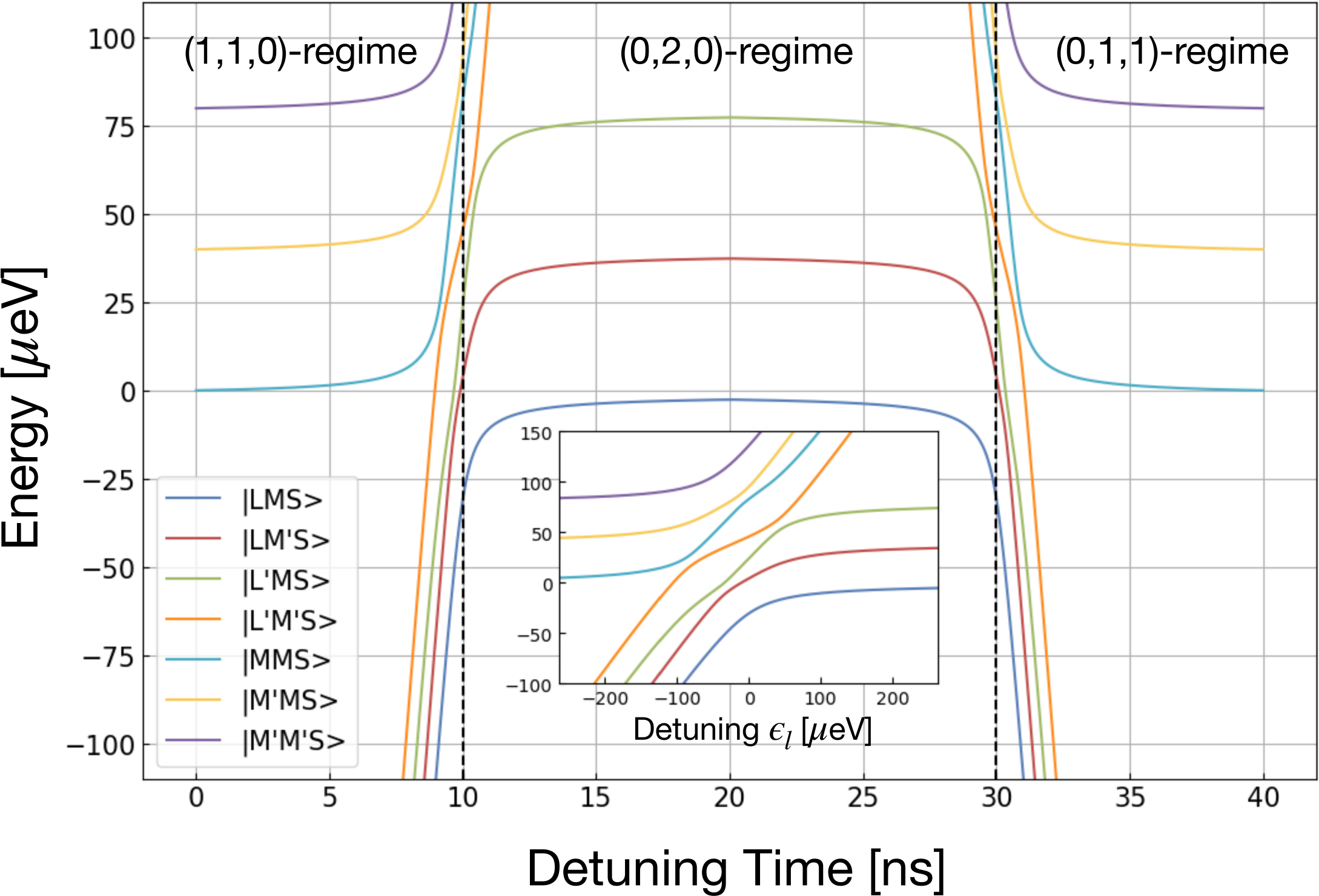}
    \caption{Energies of the relevant singlet states at every time step during the leapfrogging protocol, where the legend refers to the initial state before the procedure. The inset shows a region of $600 \, \mu \text{eV}$ centered around the avoided crossings in more detail, with labels on both axes given in units of $\mu \text{eV}$. The diagonalized Hamiltonian used the parameters from set 2, but with a middle dot valley splitting of $E_m = 40 \,\mu \text{eV}$ for easier readability.}
    \label{fig:energybands}
\end{figure}
We numerically diagonalize the Hamiltonians \eqref{eq: model,main-hamiltonian-singlet} and \eqref{eq: model,main-hamiltonian-triplet} and confirm that for adiabatic processes i.e. sufficiently slow changes in the detunings $\epsilon_i$ the ground state of the system will smoothly transition between (1,1,0)-(0,2,0)- and (0,1,1)-states. As anticipated, the triplet ground state in the (0,2,0) configuration carries a valley excitation and is therefore energetically separated from the singlet by the valley splitting. We have displayed the relevant transition and the energies of the states during the full protocol in Figure~\ref{fig:energybands}. 

\begin{table}[b]
    \caption{Sets of device parameters used for simulations.}
    \label{tab:fids}
    \begin{ruledtabular}
    \begin{tabular}{lll}
    Symbol  & Set 1 (asymmetric) & Set 2 (symmetric)
    \\ 
    \hline
    $t_l,t_l'$  & $20 , 22\, \mu \textrm{eV}$  & $18 , 18 \,\mu \textrm{eV}$ \\ 
    $t_r, t_r'$     & $16 , 19.2 \,\mu \textrm{eV}$  & $18 , 18 \,\mu \textrm{eV}$   \\ 
    $E_l, E_r$  & $50 , 40 \,\mu \textrm{eV}$ & $80 , 80\, \mu \textrm{eV}$ \\
    $E_m$  & $2 \,\mu \textrm{eV}$ & $2\, \mu \textrm{eV}$ \\
    $U_1, U_2$  & $5000 , 500 \,\mu \textrm{eV}$ & $5000 , 500 \,\mu \textrm{eV}$\\
    Detuniung speed $v$ & $100 \,\mu \textrm{eV/ns}$ & $100 \,\mu \textrm{eV/ns}$ \\
    $\left(\epsilon_{i}\right)_\text{end/start}-U _1$ & $\pm1300 \,\mu \textrm{eV}$ & $\pm1000 \,\mu \textrm{eV}$
    \end{tabular}
    \end{ruledtabular}
\end{table}
Using QuTiP \cite{qutip} to simulate the time evolution of the system, we investigate the leapfrogging process explicitly for two sets of parameters listed in Table~\ref{tab:params}. These parameter sets have been chosen so that one set represents a perfectly symmetrical TQD, whereas the other displays more asymmetric behavior while still presenting realistic device parameters found in such systems. In this way, it is possible to study whether reasonable differences between the two outer dots influence the performance of the procedure.

\begin{figure}[!h]
    \centering
    \includegraphics[width=1\linewidth]{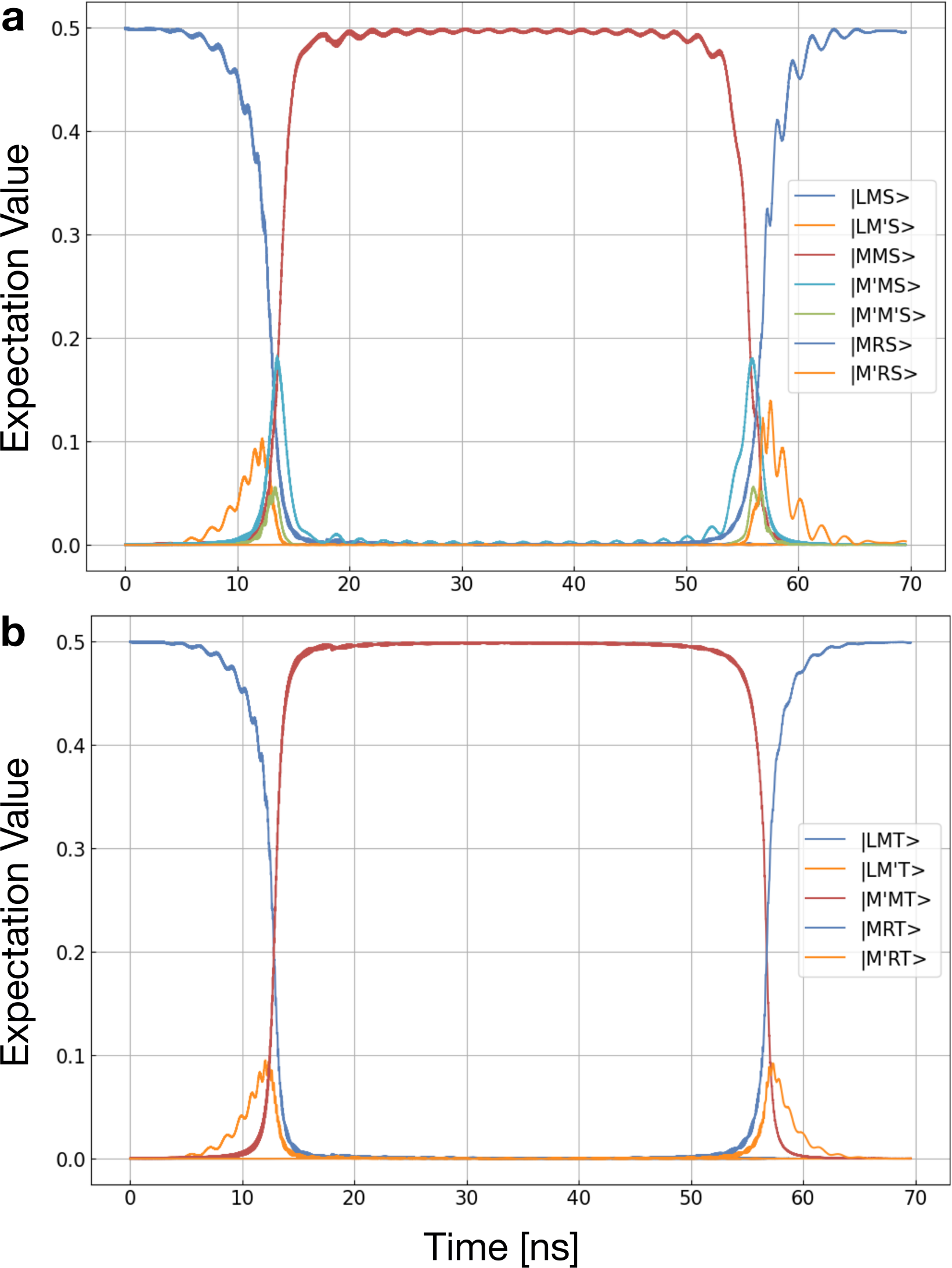}   
    \caption{Probability of finding the two-electron system in one of the basis states at each time slice for spin singlets (\textbf{a}) and triplets (\textbf{b}) simulated with parameter set 1. The system is prepared in the initial state $\ket{\psi(0)} = \frac{1}{\sqrt{2}}\left( \ket{LMT} - \ket{LMS }\right)$ and adiabatically transitions into $\ket{\psi_{\mathrm{wait}}} = \frac{1}{\sqrt{2}}\left( \ket{M'MT} - \ket{MMS }\right)$ before later transitioning into the $(0,1,1)$-regimes equivalent $\ket{\psi_{\mathrm{final}}} = \frac{1}{\sqrt{2}}\left( \ket{MRT} - \ket{MRS }\right)$.}
    \label{fig:leapfrogging1}
\end{figure}

\begin{figure}[t]
    \centering
    \includegraphics[width=1\linewidth]{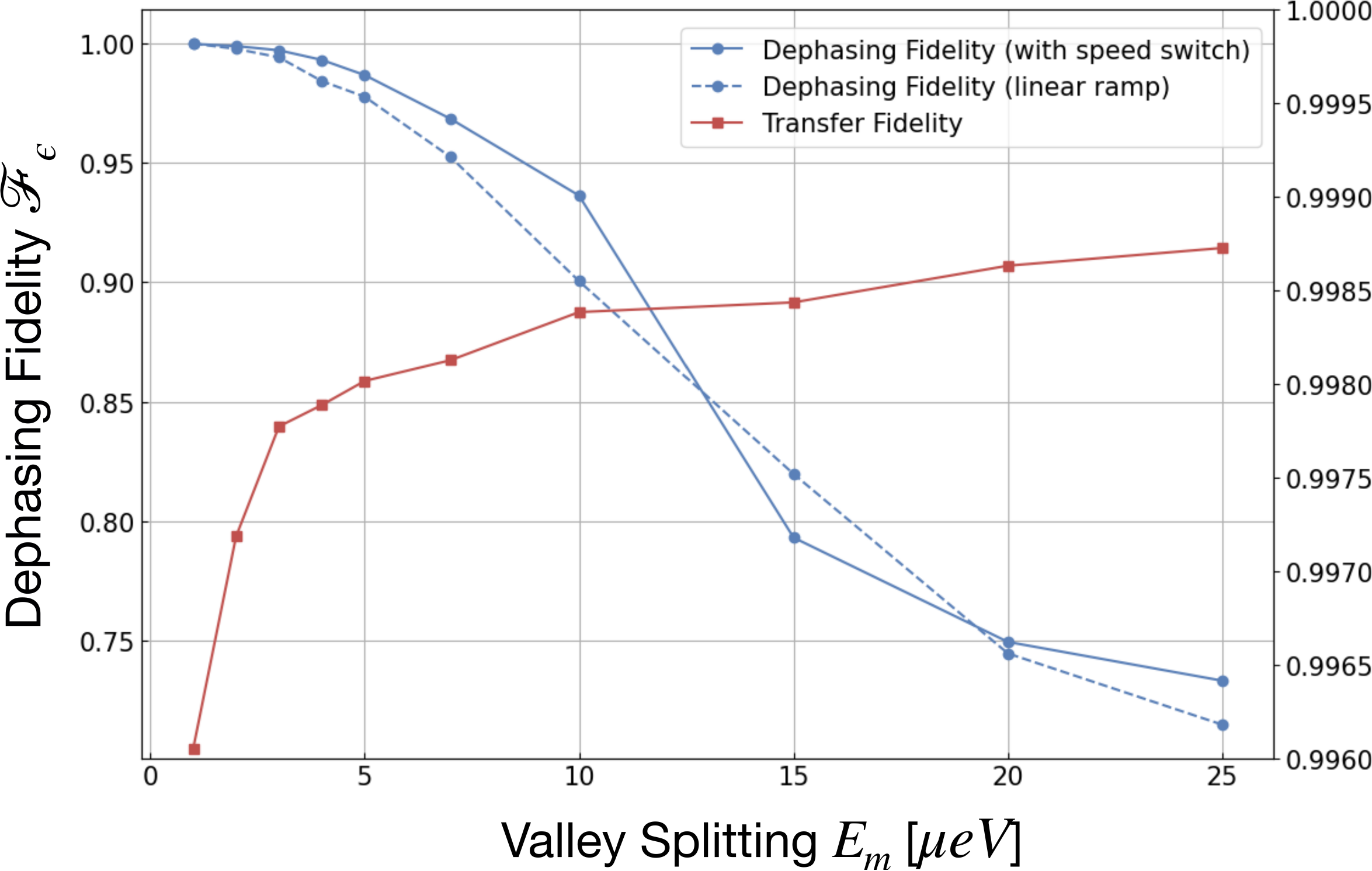} 
    \caption{Simulated fidelity associated with dephasing due to charge noise in the detunings (blue, left vertical axis) and transfer fidelity (red, right vertical axis) averaged over 100 shots with random input state and quasistatic noise realisations as a function of middle-dot valley splitting $E_m$. While the dephasing fidelity strictly drops with increasing valley splitting, the uncorrected transfer fidelity rises due to the damped ground state oscillations and plateaus around the expected error due to leakage and incomplete state transfer. The simulations were performed using parameter set 2.}
    \label{fig: fid vs valley splitting}
\end{figure}

Figure~\ref{fig:leapfrogging1} shows the expectation value of finding the system in one of the basis states at each time slice during the leapfrogging protocol. Here we used renormalized states that mix the valley excitations according to their exchange-induced coupling at the start of the detuning sequence, because they better represent the eigenstates of the system. As we perform all operations adiabatically, after the protocol is finished, the spins will return into well-defined valley states, as for strong separation these are the ground states of the systems. Therefore, one does not need to worry about partially occupying a valley-excited state during the protocol as long as one keeps track of the energies of such valley-superposition states. 
\noindent
We find that the system mostly stays in the approximated ground state at each time step and adiabatically transitions from having the mobile electron in the left to it being placed in the right dot. We also observe oscillations between the renormalized valley excited and ground states of the middle dot electron in the process of detuning the system. During the transition, we approximated the instantaneous ground states as adiabatic mixtures of the ground states at the edges of each detuning sequence. As the coupling between valley ground- and excited state is exchange-induced and depends on the detuning, this coupling, and thus the valley-state mixture in the instantaneous eigenstates of the system, changes during the protocol. Therefore the physical ground state is not captured by just the adiabatic mixture of the ground states at the beginning and the end of the sequence. As a consequence, one observes small oscillations between the ground and first excited states during the transition, which are not physical but only arise from a deviation of our basis from the actual eigenbasis of the system.
This argument is also supported by the fact that the transfer fidelity increases with increased valley splitting, as this decreases the contribution of a valley excitation in the ground state which scales with $\frac{J_c}{E_m}$, where $J_c$ is the exchange-induced coupling between the valley ground and excited states. This effect can be observed in Figure~\ref{fig: fid vs valley splitting}. 

As the simulations contain only strictly unitary dynamics we benchmark the performance of the protocol through a gate fidelity defined by the overlap of the same initial state $\ket{\psi_i}$ evolved by the ideal leapfrogging operation $\mathcal{U}$ and the imperfect noisy implementation $\tilde{\mathcal{U}}$,
\begin{equation}
    \mathcal{F} = \bra{\psi_i}\mathcal{U}^\dagger\tilde{\mathcal{U}}\ket{\psi_i}.
\end{equation}
To obtain a better understanding of the error mechanisms at play, we split the full error accounted for in the simulation into two factors. The state-transfer-fidelity $\mathcal{F}_{\mathrm{t}}$ measures leakage into noncomputational states at the end of the protocol, while the dephasing fidelity $\mathcal{F}_{\mathrm{\epsilon}}$ describes any additional errors. They are defined as follows:
\begin{gather}
        \mathcal{F}_{\mathrm{t}}\left[ \rho_f \right] = \bra{\overline{MRS}} \rho_f \ket{\overline{MRS}}+\bra{\overline{MRT}} \rho_f \ket{\overline{MRT}} \\
        \mathcal{F}_\epsilon = \frac{\mathcal{F}}{\mathcal{F}_{\mathrm{t}}}.
        \label{eq:fidelities}
\end{gather}
Here $\rho_f = \tilde{\mathcal{U}}\ket{\psi_i}\bra{\psi_i}\tilde{\mathcal{U}}^\dagger$ is the density matrix of the final state after the protocol and $\ket{\overline{MRS}}, \ket{\overline{MRT}}$ are the systems groundstates at the final detuning. 
 
Generally, we propose using TQDs with a very low valley splitting $E_m$ in the middle dot, since we find that, for linear ramp pulses, the charge transitions from one dot to another show a high sensitivity to quasistatic charge noise when the valley splitting in the middle dot is high.
This is because quasistatic noise limits the knowledge on how much time the mobile qubit spends in each dot by $\delta t = \frac{\delta \epsilon}{v}$. Since the difference in singlet-triplet splitting between the two charge configurations scales linearly with $E_m$, the additional phase collected in that unknown time interval is determined as $\delta \varphi_{\epsilon} = \frac{\delta \epsilon }{v}E_m$. Even for conservative noise levels, this will quickly lead to significant dephasing. To address this issue and reduce the severity of said effect, one can introduce a 2-level pulse sequence that switches detuning velocities during the transition. It is possible to choose the two speeds and the timing of the switch in such a way that the extra phase that is collected from transitioning earlier or later always cancels with the additional phase collected from consequently also mistiming the speed switch. Hence, this approach deterministically cancels the effect of the noise onto the phase. Assuming that the singlet and triplet ground-state energies follow those of a simple Landau-Zener transition with the same parameters, we can identify an appropriate timing and strength for the speed switch to decrease the dephasing felt during the transitions. This is still only feasible for relatively low valley splittings, as the simple Landau-Zener approximation is quite crude and decreases in accuracy with higher valley splittings. As the energy landscape is deterministic, the fidelity can therefore be significantly enhanced with more sophisticated models of the energy levels or careful calibration using available modern quantum-control techniques, consequently pushing the dephasing due to quasistatic noise in the detunings towards zero (see the Appendix~\ref{App: Charge noise during leapfrogging}). To highlight this point, we have simulated 100 instances of leapfrogging with random quasistatic noise for different middle dot valley splittings. As expected, we observe a sharp decline in fidelity with increasing valley splitting due to dephasing, as can be seen in Figure~\ref{fig: fid vs valley splitting}. We also observe that increases in infidelity do not stem from a change in transfer fidelity which rises with higher valley splittings, as this decreases the coupling between valley ground and excited states and thus suppresses the valley ground-state oscillations.  We compare this to leapfrogging without the speedswitch (dashed lines in Figure~\ref{fig: fid vs valley splitting}), where the qubit dephases significantly more quickly for most valley splittings. Another advantage of operating at low valley splittings is an additional level of charge noise protection due to $\Delta_{S-T_0}$ being essentially independent of the noisy confinement energy, as well as the valley ground and excited states sharing similar envelope functions, thereby minimizing the triplets dipole moment \cite{ErcanCoppersmith22}. In the case of higher valley splittings, there will be some mixing of the valley and orbital degree of freedom, diminishing these effects and leading to the formation of Wigner molecules \cite{ErcanCoppersmith21}.
Concerning quasistatic noise in the barrier and middle dot plunger gates, it is instructive to note that the protocol neither utilizes any change in tunnel couplings nor in the middle-dots detuning. Hence we can assume that these parameters are being held steady at only their dc bias and do not need to be quickly addressable, allowing heavy filtering of the corresponding signal and  thus greatly reducing the charge noise they are experiencing \cite{Filtered1, Filtered2}. 

An additional source of error is the misestimation of the relevant energy splitting $\Delta_{S-T_0}$ at the waiting point, which deviates from the naive middle dot valley splitting $E_m$ due to additional exchange energies. As the $S-T_0$-splitting determines the speed of the SWAP$^\gamma$-oscillations, any error in that energy will lead to timing errors of the SWAP$^\gamma$-gate and consequently negatively impact fidelity. We derive an effective formula \eqref{eq:full E_wait} for the renormalized $\Delta_{S-T_0}$ through perturbation theory. Comparing the energy splitting calculated by the formula with that obtained through numerical diagonalization, we obtain energy estimation errors of $E_{\text{wait},1} - E_{\text{num},1} \approx 1.00 \times 10^{-4}\,\mu \text{eV}$ for the first and $E_{\text{wait},2} - E_{\text{num},2} \approx 3.32 \times 10^{-5} \,\mu \text{eV}$ for the second set of parameters. Consequently, for the timescales of single oscillations, the error caused by this is negligible. For comparison, naively assuming the energy difference to be the uncorrected middle dot valley splitting $E_m$ would give energy misestimations of $E_m- E_{\text{num},1} \approx 1.53 \times 10^{-2} \,\mu \text{eV}$ and $E_m- E_{\text{num},2} \approx 1.81 \times 10^{-4} \,\mu \text{eV}$ leading to significant errors in the case of the asymmetric parameter set. 

Finally, we calculate the mean fidelities for the implementations of noisy Id-, SWAP- and $\sqrt{\text{SWAP}}$-gates. To achieve this, we first simulate the noiseless leapfrogging process for an arbitrary pure quantum state to obtain the deterministic detuning phases for the corresponding parameter regime. As these phases are independent of the input state, we can calculate them once for each of the sets of parameters and then compensate for these phases by waiting an additional offset time $t_{o}$ in every implementation of the leapfrogging, thereby implementing an Id-gate by default. Next we utilize the Fubini-Study metric to generate 1000 random pure two-electron states initialized in the (1,1,0)-charge configuration, both electrons occupying the valley ground state. As all polarized triplets will experience practically the same Hamiltonian evolution as the $T_0$-triplet up to some deterministic phase difference, we project the triplet components of each state onto the $\ket{T_0}$ component, thereby easing the computational effort required for the simulation while still accounting for the polarized triplets in the sampling. For each random input state, we generate an instance of quasistatic noise drawn from a zero-mean Gaussian distribution with standard deviation $\langle \delta \epsilon^2 \rangle = 4.4 \,\mu \text{eV}$~\cite{Connors2022-ik} and simulate an instance of the Id-, SWAP-, and $\sqrt{\text{SWAP}}$-gate for each manifestation of the noise.

\begin{table}[h!]
    \caption{Error budget: Error probabilities from various sources, and total error probability for the two previously defined parameter sets.}
    \label{tab:params}
    \begin{ruledtabular}
    \begin{tabular}{lll}
    Error Source   & Set 1  & Set 2
    \\ 
    \hline
    State Transfer  & $1.70\times 10^{-3} $  & $1.42 \times 10^{-3} $  \\ 
    Detuning Noise & $1.69 \times 10^{-3}$ & $1.14 \times 10^{-3}$   \\
    \textcolor{mygray}{Detuning noise (no switch)}& \textcolor{mygray}{$1.69 \times 10^{-3}$} & \textcolor{mygray}{$2.31 \times 10^{-3}$}   \\
    Tunnelling Noise & $1.07 \times 10^{-3}$& $1.65\times 10^{-3}$  \\
    \hline
    Total error probability    & $4.45\times 10^{-3}$  & $4.20 \times 10^{-3}$   
    \end{tabular}
    \end{ruledtabular}
\end{table} 
As mentioned earlier, there is a mathematical mismatch in the estimated groundstates used in the calulations and those found in the actual system leading to artificial oscillations in the state transfer fidelity.
 To minimize the virtual error due to these oscillations, we consider the leakage of the operation for different wait times and then search for a maximum in the resulting oscillations while keeping the detuning range constant. This is justified, as physically the waiting process only leads to a phase accumulation of the triplet states and does not influence the leakage  whatsoever. Therefore the value at such maximum should reflect the actual state transfer fidelity.
Following this procedure we obtain average transfer infidelities of $1-\mathcal{F}_1 \approx 1.02 \times 10^{-3}$ for the first and $1-\mathcal{F}_2 \approx 1.45 \times 10^{-3}$ for the second set of parameters which are listed in Table~\ref{tab:fids}. For both sets of parameters the difference in Zeeman energy $E_Z$ and middle-dot valley splitting $E_V$ is also high enough such that there is no crossing or sufficiently close approaching of the $\ket{S}$- and $\ket{T_-}$-states energy, to provoke a spin-orbit-coupling induced $S-T_-$ anticrossing, whereby leakage from such channel can safely be neglected (see Appendix~\ref{App:SO-Grad}). 

The dephasing mechanisms explicitely included in the simulations are the quasistatic noise in the left and right dot detunings as well as the misestimation of the energy splitting at the waiting point. We extract the dephasing error $1-\mathcal{F}_\epsilon$ from the full fidelity by calculating the explicit transfer fidelity of each instance of a simulated SWAP$^\gamma$-gate and dividing the full fidelity by that value as defined in Eq.~\eqref{eq:fidelities}. The resulting dephasing infidelities of $1-\mathcal{F}_{\epsilon,1} \approx 1.69 \times 10^{-3}$ for the first and $1-\mathcal{F}_{\epsilon,2} \approx 1.14 \times 10^{-3}$ for the second set then stem fully from the aforementioned dephasing errors. We find that, as anticipated, the dephasing error due to energy misestimation and detuning noise depends on the wait time to only a miniscule degree. The differences in dephasing error between the three different implementations of the SWAP$^\gamma$ -gate is on the order of $\sim 10^{-5}$, which agrees with the numerically estimated deviation. As it is  deterministic, it could even be reduced further by careful calibration and control electronics with sufficiently high sampling rates in a realistic setting. We also see that the speed switch makes a significant difference in fidelity only for the second set of parameters. The reason for this most likely the asymmetry in the first set leading to a stronger deviation of the energies from the naive LZ- model. Still, this demonstrates that the speed switch method can cancel out the dephasing due to quasistatic noise in the detuning parameters and that the given formula for the speed switch serves as a good starting point for further calibration and fine tuning. 

To account for the error induced by charge noise in the tunneling parameter, we can take the product of the leakage and dephasing errors obtained from the simulations and multiply it by the imperfect fidelity expected to occur due to tunneling errors derived in Appendix~\ref{App: Charge noise during leapfrogging}. Because the tunnel barrier is not required to be quickly addressable we can expect a significantly lower charge noise amplitude of $2 \langle \delta t_c\rangle = 0.05 \, \mu \text{eV}$ leading us to expect dephasing errors from quasistatic noise in the tunneling on the scale of $1-\mathcal{F}_{t_c,1} \approx 1.07 \times 10^{-3}$ and $1-\mathcal{F}_{t_c,2} \approx 1.65 \times 10^{-3}$ for both sets of parameters. 

Combining all these error mechanisms, our considerations lead us to expect gate infidelities of $1-\mathcal{F}_{\text{Id},1} \approx 4.45\times 10^{-3}$ and $1-\mathcal{F}_{\text{Id},2} \approx 4.20 \times 10^{-3}$ for the identity gates for the respective parameter sets, with the different powers of the SWAP$^\gamma$ gates expected to perform similarly well. The gate times of the leapfrogging tuned to performing the identity on the mobile qubit then amount to $t_{\mathrm{final},1} \approx 69.76$~ns and $t_{\mathrm{final},2} \approx 48.66$~ns, making this operation comparable in speed to regular two-qubit gates in similar devices \cite{undseth2026weightfourparitycheckssilicon}. 

These numbers could be improved further by applying more involved methods for identifying the optimal speedswitch parameters (or performing careful calibration in experiment), more sophisticated pulse shaping and better control electronics leading to reduced charge noise. Taking into account these facts, the protocol promises to meet the error-correction threshold by an even greater margin for both sets of parameters. It is important to note that some of the error arising from leakage is intrinsic to the operation and can only partially be reduced without opening the door to other error mechanisms. For example, one could decrease the leakage from diabatic processes by detuning more slowly, in exchange increasing the charge noise sensibility and prolonging the gate time. The asymmetric parameter set in the cases considered performs similarly well as the symmetric one, revealing only a limited sensitivity to deviation between the outer dots. In cases of higher charge noise amplitudes in the tunneling, one expects the asymmetric set to perform better, because the error due to tunneling noise strongly favors setups for which $\alpha = t_i' / (\sqrt{2}t_i )\approx 1$. In the case of a symmetric sweep, satisfying this condition strictly leads to a complete cancelation of any tunneling noise.

\section{Conclusions}
\label{Conclusions}
We investigated the dynamics and theoretical feasibility of bucket-brigade shuttling an electron through an already occupied quantum dot. While we found that for the current level of achievable pulse control, the central dot has to lie in a region of low valley splitting to offer reasonably low decoherence levels, we also demonstrated that in this particular case, leapfrogging can be implemented as a well-behaved quantum operation. By simulating the protocol for two different sets of realistic device parameters, we see not only that the spin can be reliably transferred through the occupied dot, but that the resulting SWAP$^\gamma$ gate offers a new method for entangling a mobile and stationary qubit with high fidelity. Although it is challenging to predict the phase collected during the detuning processes because it strongly depends on the chosen parameters, the phase is resilient enough against typical levels of quasistatic noise in the case where the valley splitting in the middle dot is small. One can therefore map the collected phase once for given TQD parameters by preparing the two spins in a predetermined product state of different spins and then measure the orientation of the mobile qubit after leapfrogging, e.g., using Pauli spin blockade readout. By doing this for different waiting times, one obtains the deterministic detuning phases and can thus compensate for them by modifying the waiting time in the $(0,2,0)$ configuration. Whilst not discussed in full detail, longitudinal magnetic field gradients along the TQD add a partial CPHASE-gate character to the leapfrogging operation. Although these can lead to detrimental reductions in gate-fidelity when unaccounted for, tuning the pulses accordingly can compensate for this deterministic effect fairly easily. In particular, for strong gradients, this opens up an additional mode of operation for leapfrogging, functioning in an analogue manner to the one presented here, but implementing a CZ-phase gate on the involved qubits instead as hinted in Appendix~\ref{App:SO-Grad}.

In this paper, we did not investigate the relaxation- and decoherence times of the $(0,2,0)$-state, though we see no reason for any significant increase in spin or valley relaxation times, which are typically listed to be around $T_{1,\text{spin}} \sim 1 \, {\rm s}$ \cite{spinrelaxation} and $T_{1,\text{valley} } \approx \, 10 \,{\rm ms}$ \cite{valleyrelaxation} and thus by themselves long enough to not matter for the timescales considered here (provided that one operates sufficiently far away from any spin-valley hotspot). A less discussed mechanism for decoherence might arise from the utilization of a superposition of the valley ground and excited states in the protocol. Since the valley splitting is varying rapidly throughout typical semiconductor quantum processor chips \cite{valleycorrelations}, any charge-noise-induced fluctuation in the middle dot position translates to similar fluctuations of the valley splitting energy, thereby dephasing the singlet-triplet superposition. While this requires further investigation, the error caused by this effect should be limited \cite{spectatorqubit}, in particular as both the barrier and middle-dot plunger gates do not need to be addressable with fast pulses, consequently limiting their noise and the resulting fluctuations in the dot position and valley splitting. Furthermore, when operating near a valley splitting minimum, which is proposed to be the most impactful and suitable application of leapfrogging, the protocol obtains another level of first-order charge-noise protection. One can also expect a significant increase of the fidelity from more sophisticated pulse-shape optimization techniques \cite{OptimisedPulse,GRAPE}. Anticipating improvements in noise levels and precision of control electronics, one could also envision applying leapfrogging to regions of higher valley splitting or utilizing different kinds of energetically low-lying excitations instead. While the associated increased difficulty in pulse timing and quasistatic noise would be handled by the electronics, one would additionally lose some charge noise protection and gain a potentially harmful electric dipole moment due to the nature of the orbital single-dot two-electron wave functions \cite{ErcanCoppersmith21}. Therefore, this operational mode requires more careful control and further study to asses its potential feasibility. 

Generally speaking, leapfrogging can enable an additional operation to implement entangling gates in future architectures.  Moreover, it may provide an interface for a chip architecture with a  ``highway'' layer of mobile shuttled electrons transferring information between smaller stationary qubit array cells based on information transfer employing SWAP operations. Leapfrogging could allow for an easier and more practicable path to reordering qubits running in the same conveyor belt network, thereby simplifying the scheduling problem in the implementation of quantum circuits. Thus, this operation offers a path to a more efficient use of the already established and functional method of spin shuttling\cite{undseth2026weightfourparitycheckssilicon,SiegelBenjamin2026Quantum}. First and foremost, however, leapfrogging provides a practical way to isolate and exploit otherwise dangerous low-valley-splitting regions of a silicon qubit device. The fabrication of a semiconductor device capable of verifying the fidings presented in this work appears possible  with current state-of-the-art technology, allowing for initial tests and implementations to be performed in the near future.

\section{Acknowledgements}
\label{Acknowledgements}
We acknowledge financial support by the European Union through the Horizon Europe project QLSI2.

\appendix
\section{The Hamiltonian}
\label{App: The Hamiltonian}
In an effort to derive a full Hamiltonian for the two-electron triple quantum dot including the valley degree of freedom, we first start by constructing a bare one-electron spinless Hamiltonian. For this purpose, we assume that the direct interdot tunnel coupling $t_o$ between the outer (left and right) dots is vanishingly small, such that the Hamiltonian can be written as,
\begin{equation}
    \begin{aligned}
         H_{\text{bare}} = &\sum_{i \in \left\{l,m,r \right\}}  \sum_{v \in \left\{ +,- \right\}} \epsilon_i  c_{i,v}^\dagger c_{i,v} \\
                        + &\sum_v (t_c)_l \left( c_{l,v}^\dagger c_{m,v} + c_{m,v}^\dagger c_{l,v} \right) \\
                        + &\sum_v (t_c)_r \left( c_{m,v}^\dagger c_{r,v} + c_{r,v}^\dagger c_{m,v} \right) \\
                        + &\sum_{i} \lvert \Delta_i \rvert \left( e^{i\phi_i} c_{i,-}^\dagger c_{i,+} + e^{-i\phi_i}c_{i,+}^\dagger c_{i,-} \right),
    \end{aligned}
\end{equation}  
where $(t_c)_{l,r}$ refers to the tunnel coupling between the left/right and the middle quantum dot which can be chosen to be real, $c_{i,v}$ is the annihilation operator of the electron in the $i$th dot in the $v \in \left\{ +,- \right\}$ valley state. Throughout this paper, $+$,$-$ refer to the $\ket{+k_0}$ and $\ket{-k_0}$ valley states rather than the valley ground and excited states as sometimes encountered in the literature. As the left and the right dot are not directly coupled, we are allowed to rotate the system in such a way that the valley phase of the middle dot is zero and we only need to care about the valley phase differences $\theta_{l,r} = \phi_m - \phi_{l,r}$ between the middle and outer dots \cite{valley-hamiltonian}. Therefore, we arrive at the following bare Hamiltonian,
\begin{equation}
        \begin{pmatrix}
            \epsilon_l & 0  & 0 & 0 & t_l & t'_l\\
            0 & \epsilon_l+E_l & 0 & 0 & -t'_l & t_l \\
            0 & 0 & \epsilon_r & 0 & t_r &  t'_r \\
            0& 0 & 0 & \epsilon_r + E_r & -t'_r & t_r  \\
            t_l & -t'_l & t_r & -t'_r & 0  & 0 \\
            t'_l & t_l & t'_r & t_r & 0 &   E_m  
        \end{pmatrix},
    \label{eq: App1 bare-valley-h}
\end{equation}
which is written in the basis
\begin{gather*}
    \hspace{5mm} \left\{ \ket{L}, \ket{L'}, \ket{R}, \ket{R'}, \ket{M}, \ket{M'} \right\},
    \end{gather*}
and where we have used the notations,
\begin{gather*}
        t_i = (t_c)_i \cos \left( \frac{\theta_i}{2} \right), 
        \hspace{5mm} 
        t'_i = (t_c)_i \sin \left(  \frac{\theta_i}{2}\right).
\end{gather*}
Here we absorbed the detuning $\epsilon_m$ of the middle dot into the detunings of the outer dots and aligned the global valley-quantization axis with the valley quantization axis of the middle dot. Setting aside spin-orbit coupling and assuming vanishing magnetic field gradients $\Delta B_\parallel$ and $\Delta B_\perp$, adding the spin degree of freedom will give us two uncoupled copies of the same Hamiltonian~\eqref{eq: App1 bare-valley-h} with diagonal elements shifted by the Zeeman energy $E_z^{(i)}$ of the respective dot for each spin state.
Adding a second electron to this setup, the system is characterized by the following Hamiltonian,
\begin{equation}
    \begin{aligned}
        H &= \sum_{i}  \Bigg( U_1^{(i)} \hat{n}_i(\hat{n}_i-1) 
            + \sum_{j< i} U_2^{(i,j)} \hat{n}_i \hat{n}_j +  \epsilon_i \hat{n}_i \\
            &+\sum_{v \in \left\{+,- \right\}}  \frac{E_z^{(i)}}{2} \left(c_{i,\uparrow,v}^\dagger c_{i,\uparrow,v} - c_{i,\downarrow,v}^\dagger c_{i,\downarrow,v}\right) \Bigg)\\
            &+ \sum_{i,s} \lvert\Delta_i\rvert\left( e^{i\phi_i} c_{i,s,-}^\dagger c_{i,s,+} + e^{-i\phi_i}c_{i,s,+}^\dagger c_{i,s,-} \right) \\
            &+ \sum_{s,v} (t_c)_l \left( c_{l,s,v}^\dagger c_{m,s,v} + c_{m,s,v}^\dagger c_{l,s,v} \right) \\
            &+ \sum_{s,v} (t_c)_r \left( c_{m,s,v}^\dagger c_{r,s,v} + c_{r,s,v}^\dagger c_{m,s,v} \right) .
    \end{aligned}
\end{equation}
This Hamiltonian contains several charge states that are irrelevant for the proposed operation. We project out these charge states using a Schrieffer-Wolff transformation, resulting in an effective Hamiltonian which  again does not contain any coupling between the different spin states. This suggests the choice of the singlet-triplet basis in which the Hamiltonian decomposes into four independent blocks, $H_S$ for the singlet and $H_{T,s}$ with $s \in \left\{ 0, \pm \right\}$ for the three
triplets. Performing again the same valley rotation described in~\eqref{eq: App1 bare-valley-h}, the Hamiltonians for the triplets read as,
\begin{widetext}
    \begin{equation}
        \label{eq: model,main-hamiltonian-triplet}
        \hat{H}_{T,s} =  
        \begin{pmatrix}
            E_l + \epsilon_l & 0 & 0 & 0 & 0 & 0 & 0 & 0 & t_l\\0 & E_l + E_m + \epsilon_l & 0 & 0 & 0 & 0 & 0 & 0 & t_l'\\0 & 0 & E_r + \epsilon_r & 0 & 0 & 0 & 0 & 0 & - t_r\\0 & 0 & 0 & E_m + E_r + \epsilon_r & 0 & 0 & 0 & 0 & - t_r'\\0 & 0 & 0 & 0 & \epsilon_l & 0 & 0 & 0 & t_l'\\0 & 0 & 0 & 0 & 0 & E_m + \epsilon_l & 0 & 0 & - t_l\\0 & 0 & 0 & 0 & 0 & 0 & \epsilon_r & 0 & - t_r'\\0 & 0 & 0 & 0 & 0 & 0 & 0 & E_m + \epsilon_r & t_r\\t_l & t_l' & - t_r & - t_r' & t_l' & - t_l & - t_r' & t_r & E_m + U_1
        \end{pmatrix}
        + E_s,
    \end{equation}
    in the basis,
    \begin{equation*}
  \hspace{5mm} \left\{ \ket{L'MT_s}, \ket{L'M'T_s}, \ket{MR'T_s}, \ket{M'R'T_s},\ket{LMT_s}, \ket{LM'T_s}, \ket{MRT_s}, \ket{M'RT_s}, \ket{M'MT_s} \right\}.
    \end{equation*}
\end{widetext}
Here we omitted the additional exchange couplings resulting from the redundant charge states, as they all have coupling strengths $< 0.1 \,\mu eV$ and should as a result lead to no significant additional dynamics in the protocol. The only terms added through this procedure that meet this threshold are the diagonal exchange terms given by,
\begin{widetext}
    \begin{equation*}
        \begin{aligned}
            J_{1}^d &= 
\frac{t_l^{2}}{- U_{1} - \epsilon_l} + \frac{t_r^{2}}{U_{2} - \epsilon_r} + \frac{t_r'^{2}}{- E_r + U_{2} - \epsilon_r} \approx -\frac{t_l^2}{\epsilon_l+U_1} + \frac{(t_c)_r^2}{-\epsilon_r+U_2},
\\
J_{2}^d &= 
\frac{t_l'^{2}}{E_m - U_{1} - \epsilon_l} + \frac{t_r^{2}}{E_m - E_r + U_{2} - \epsilon_r} + \frac{t_r'^{2}}{E_m + U_{2} - \epsilon_r} \approx -\frac{t_l'^2}{\epsilon_l+U_1} + \frac{(t_c)_r^2}{-\epsilon_r+U_2}
,\\
J_{3}^d &= 
\frac{t_l^{2}}{U_{2} - \epsilon_l} + \frac{t_l'^{2}}{- E_l + U_{2} - \epsilon_l} + \frac{t_r^{2}}{- U_{1} - \epsilon_r} \approx -\frac{t_r^2}{\epsilon_r+U_1} + \frac{(t_c)_l^2}{-\epsilon_l+U_2}
,\\
J_{4}^d &= 
\frac{t_l^{2}}{- E_l + E_m + U_{2} - \epsilon_l} + \frac{t_l'^{2}}{E_m + U_{2} - \epsilon_l} + \frac{t_r'^{2}}{E_m - U_{1} - \epsilon_r}\approx -\frac{t_r'^2}{\epsilon_r+U_1} + \frac{(t_c)_l^2}{-\epsilon_l+U_2}
,\\
J_{5}^d &= 
\frac{t_l'^{2}}{- E_l - U_{1} - \epsilon_l} + \frac{t_r^{2}}{U_{2} - \epsilon_r} + \frac{t_r'^{2}}{- E_r + U_{2} - \epsilon_r} \approx -\frac{t_l'^2}{\epsilon_l+U_1} + \frac{(t_c)_r^2}{-\epsilon_r+U_2} \approx J_2^d
,\\
J_{6}^d &= 
\frac{t_l^{2}}{- E_l + E_m - U_{1} - \epsilon_l} + \frac{t_r^{2}}{E_m - E_r + U_{2} - \epsilon_r} + \frac{t_r'^{2}}{E_m + U_{2} - \epsilon_r}\approx -\frac{t_l^2}{\epsilon_l+U_1} + \frac{(t_c)_r^2}{-\epsilon_r+U_2} \approx J_1^d
,\\
J_{7}^d &= 
\frac{t_l^{2}}{U_{2} - \epsilon_l} + \frac{t_l'^{2}}{- E_l + U_{2} - \epsilon_l} + \frac{t_r'^{2}}{- E_r - U_{1} - \epsilon_r}\approx -\frac{t_r'^2}{\epsilon_r+U_1} + \frac{(t_c)_l^2}{-\epsilon_l+U_2} \approx J_4^d
,\\
J_{8}^d &= 
\frac{t_l^{2}}{- E_l + E_m + U_{2} - \epsilon_l} + \frac{t_l'^{2}}{E_m + U_{2} - \epsilon_l} + \frac{t_r^{2}}{E_m - E_r - U_{1} - \epsilon_r}\approx -\frac{t_r^2}{\epsilon_r+U_1} + \frac{(t_c)_l^2}{-\epsilon_l+U_2} \approx J_3^d
,\\
J_{9}^d &= 
0
.
        \end{aligned}
    \end{equation*}
\end{widetext}
Again, plugging in typical parameters, we find that the diagonal terms are all similar in magnitude. This is to be expected, as the most significant term is the bare interdot coupling, independent of the valley phase, which one would always tune to be of similar strength, while the first valley-phase dependent term is more heavily damped by the $\epsilon_i + U_1$ denominator. As a result, all states will experience a similar shift in energy which can consequently be absorbed into the detunings, especially considering the fact that the difference between the terms inside the same charge sector will always only be of strength $(t_i^2-t_i'^2)/(\epsilon_i+U_1)$ which is always very small.
Similarly, the matrix representing the spin-singlet Hamiltonian $H_S$ is calculated as follows,
\begin{widetext}
    \begin{gather}
            \label{eq: model,main-hamiltonian-singlet}
            \hat{H}_S =  \nonumber\\
            \begin{pmatrix}
                E_l + \epsilon_l & 0 & 0 & 0 & 0 & 0 & 0 & 0 & t_l & - \sqrt{2} t_l' & 0 \\
                0 & E_l + E_m + \epsilon_l & 0 & 0 & 0 & 0 & 0 & 0 & - t_l' & 0 & \sqrt{2} t_l\\
                0 & 0 & E_r + \epsilon_r & 0 & 0 & 0 & 0 & 0 & t_r & - \sqrt{2} t_r' & 0\\
                0 & 0 & 0 & E_m + E_r + \epsilon_r & 0 & 0 & 0 & 0 & - t_r' & 0 & \sqrt{2} t_r\\
                0 & 0 & 0 & 0 & \epsilon_l & 0 & 0 & 0 & t_l' & \sqrt{2} t_l & 0\\
                0 & 0 & 0 & 0 & 0 & E_m + \epsilon_l & 0 & 0 & t_l & 0 & \sqrt{2} t_l'\\
                0 & 0 & 0 & 0 & 0 & 0 & \epsilon_r & 0 & t_r' & \sqrt{2} t_r & 0\\
                0 & 0 & 0 & 0 & 0 & 0 & 0 & E_m + \epsilon_r & t_r & 0 & \sqrt{2} t_r'\\
                t_l & - t_l' & t_r & - t_r' & t_l' & t_l & t_r' & t_r & E_m + U_1& 0 & 0\\
                - \sqrt{2} t_l' & 0 & - \sqrt{2} t_r' & 0 & \sqrt{2} t_l & 0 & \sqrt{2} t_r & 0 & 0 & U_1& 0\\
                0 & \sqrt{2} t_l & 0 & \sqrt{2} t_r & 0 & \sqrt{2} t_l' & 0 & \sqrt{2} t_r' & 0 & 0 & 2 E_m + U_1 
            \end{pmatrix},
    \end{gather}
    where we have used the basis,
    \begin{equation*}
 \hspace{5mm} \left\{ \ket{L'MS}, \ket{L'M'S}, \ket{MR'S}, \ket{M'R'S}, \ket{LMS}, \ket{LM'S}, \ket{MRS}, \ket{M'RS}, \ket{M'MS}, \ket{MMS}, \ket{M'M'S} \right\}.
    \end{equation*}
\end{widetext} 
We are interested in setups where the incoming electron carries no valley excitation and where the valley splittings of the outer dots are significantly larger than that of the middle dot. Therefore, states with outer valley excitations are energetically separated from the others and can also be gotten rid of by another Schrieffer-Wolff transformation leading us to the following Hamiltonian,
\begin{widetext}
    \begin{equation}
    \label{eq:Full-Exchange-Hamiltonian}
        H_{S} = \begin{pmatrix}
            \epsilon_l+J_1^d & J_{l}  & J_{13} & J_{14} & t_l' & \sqrt{2} t_l & 0\\
            J_{l} & \epsilon_l+E_m +J_2^d& J_{23} & J_{24} & -t_l & 0 & -\sqrt{2} t'_l\\
            J_{13} & J_{23} & \epsilon_r +J_3^d& J_{r} & -t'_r & -\sqrt{2} t'_r & 0\\
            J_{14} & J_{24} & J_{r} & \epsilon_r + E_m+J_4^d& t_r & 0  & \sqrt{2} t'_r\\
            t'_l & -t_l & -t'_r & t_r & U_1+E_m +J_5^d & J_m & J_m \\
            \sqrt{2} t_l & 0 & -\sqrt{2} t'_r & t_r & J_m &   U_1+J_6^d & 0\\
            0 & -\sqrt{2}t'_l &  0 & \sqrt{2} t'_r & J_m &   0  & U_1+2 E_m+J_7^d
        \end{pmatrix},
    \end{equation}
with the definitions,
    \begin{equation}
        \begin{aligned}
            J_{l} &= 
- \frac{t_l t_l'}{- 2 E_l + 2 E_m - 2 U_1- 2 \epsilon_l} + \frac{t_l t_l'}{E_m - U_1- \epsilon_l} - \frac{t_l t_l'}{- 2 E_l - 2 U_1- 2 \epsilon_l} - \frac{t_r t_r'}{2 E_m - 2 E_r + 2 U_{2} - 2 \epsilon_r}, \\ 
&+ \frac{t_l t_l'}{- U_1- \epsilon_l}- \frac{t_r t_r'}{- 2 E_r + 2 U_{2} - 2 \epsilon_r} + \frac{t_r t_r'}{2 E_m + 2 U_{2} - 2 \epsilon_r} + \frac{t_r t_r'}{2 U_{2} - 2 \epsilon_r} \approx -\frac{t_l t_l'}{U_1+\epsilon_l},
\\
J_{13} &= 
\frac{t_l t_r}{2 U_{2} - 2 \epsilon_r} + \frac{t_l t_r}{2 U_{2} - 2 \epsilon_l},
\\
J_{14} &= 
\frac{t_l' t_r}{2 E_m + 2 U_{2} - 2 \epsilon_l} + \frac{t_l' t_r}{2 U_{2} - 2 \epsilon_r},
\\
J_{23} &= 
\frac{t_l t_r'}{2 E_m + 2 U_{2} - 2 \epsilon_r} + \frac{t_l t_r'}{2 U_{2} - 2 \epsilon_l},
\\
J_{24} &= 
\frac{t_l' t_r'}{2 E_m + 2 U_{2} - 2 \epsilon_r} + \frac{t_l' t_r'}{2 E_m + 2 U_{2} - 2 \epsilon_l},
\\
J_{r} &= 
- \frac{t_l t_l'}{- 2 E_l + 2 E_m + 2 U_{2} - 2 \epsilon_l} + \frac{t_l t_l'}{2 E_m + 2 U_{2} - 2 \epsilon_l} - \frac{t_l t_l'}{- 2 E_l + 2 U_{2} - 2 \epsilon_l} + \frac{t_l t_l'}{2 U_{2} - 2 \epsilon_l}, \\ 
&- \frac{t_r t_r'}{2 E_m - 2 E_r - 2 U_1- 2 \epsilon_r} - \frac{t_r t_r'}{- 2 E_r - 2 U_1- 2 \epsilon_r} + \frac{t_r t_r'}{E_m - U_1- \epsilon_r} + \frac{t_r t_r'}{- U_1- \epsilon_r} \approx  -\frac{t_r t_r'}{U_1+\epsilon_r},
\\
J_{m} &= 
- \frac{\sqrt{2} t_l t_l'}{- 2 E_l + 2 E_m + 2 U_1- 2 \epsilon_l} - \frac{\sqrt{2} t_l t_l'}{- 2 E_l + 2 U_1- 2 \epsilon_l} - \frac{\sqrt{2} t_r t_r'}{2 E_m - 2 E_r + 2 U_1- 2 \epsilon_r} - \frac{\sqrt{2} t_r t_r'}{- 2 E_r + 2 U_1- 2 \epsilon_r}\\
& \approx \frac{\sqrt{2} t_l t_l'}{ E_l - U_1+ \epsilon_l} + \frac{\sqrt{2} t_r t_r'}{ E_r - U_1+ \epsilon_r}, \\ 
J_{1}^d &= 
\frac{2 t_l^{2}}{- U_1- \epsilon_l} + \frac{t_l'^{2}}{- E_l - U_1- \epsilon_l} + \frac{t_r^{2}}{U_{2} - \epsilon_r} + \frac{t_r'^{2}}{- E_r + U_{2} - \epsilon_r},
\\
J_{2}^d &= 
\frac{t_l^{2}}{- E_l + E_m - U_1- \epsilon_l} + \frac{2 t_l'^{2}}{E_m - U_1- \epsilon_l} + \frac{t_r^{2}}{E_m - E_r + U_{2} - \epsilon_r} + \frac{t_r'^{2}}{E_m + U_{2} - \epsilon_r},
\\
J_{3}^d &= 
\frac{t_l^{2}}{U_{2} - \epsilon_l} + \frac{t_l'^{2}}{- E_l + U_{2} - \epsilon_l} + \frac{2 t_r^{2}}{- U_1- \epsilon_r} + \frac{t_r'^{2}}{- E_r - U_1- \epsilon_r},
\\
J_{4}^d &= 
\frac{t_l^{2}}{- E_l + E_m + U_{2} - \epsilon_l} + \frac{t_l'^{2}}{E_m + U_{2} - \epsilon_l} + \frac{t_r^{2}}{E_m - E_r - U_1- \epsilon_r} + \frac{2 t_r'^{2}}{E_m - U_1- \epsilon_r},
\\
J_{5}^d &= 
\frac{t_l^{2}}{- E_l + E_m + U_1- \epsilon_l} + \frac{t_l'^{2}}{- E_l + U_1- \epsilon_l} + \frac{t_r^{2}}{E_m - E_r + U_1- \epsilon_r} + \frac{t_r'^{2}}{- E_r + U_1- \epsilon_r},
\\
J_{6}^d &= 
\frac{2 t_l'^{2}}{- E_l + U_1- \epsilon_l} + \frac{2 t_r'^{2}}{- E_r + U_1- \epsilon_r},
\\
J_{7}^d &= 
\frac{2 t_l^{2}}{- E_l + E_m + U_1- \epsilon_l} + \frac{2 t_r^{2}}{E_m - E_r + U_1- \epsilon_r}.
    \end{aligned}
    \label{eq: Singlet exchange terms}
    \end{equation}
\end{widetext}
The different magnitudes of the parameters $E_m \ll E_l$, $E_r < U_2 \ll U_1, \epsilon_l, \epsilon_r$ allow us to  simplify the Hamiltonian in order to optimize its readability. In the numerical calculations that follow, the original formulas will be applied. We see now that the three $(0,2,0)$ states are  separated by exactly $E_m^m = E_m+\frac{1}{2}\left( J_7^d - J_6^d \right)$. Since $\epsilon_l,\epsilon_r \approx U_1$ and $\epsilon_i-U_2 \approx \epsilon_i$ during the protocol, we can ignore the small change in the diagonal exchange terms during the leapfrogging and absorb them into the definitions of $\epsilon_i' = \epsilon_i + J_i^d - J_6^d$ and renormalized valley excitations $E_m^i = E_m + J_{i+1}^d-J_i^d$. Additionally, we can ignore the effect of the small couplings $J_{13},J_{14},J_{23},J_{24} \sim 10^{-2} \mu eV$ between the (1,1,0)- and (0,1,1)-states, as whenever these states are occupied to a meaningful degree, their energy difference will be orders of magnitude larger than the coupling. These considerations let us simplify the Hamiltonian~\eqref{eq:Full-Exchange-Hamiltonian} into the following form,
\begin{widetext}
    \begin{equation}
        H_{S} = \begin{pmatrix}
            \epsilon_l'& J_{l}  & 0 & 0 & t_l' & \sqrt{2} t_l & 0\\
            J_{l} &  \epsilon_l'+E_m^l & 0 & 0 & t_l & 0 & \sqrt{2} t'_l\\
            0 & 0 &  \epsilon_r' & J_{r} & t'_r & \sqrt{2} t'_r & 0\\
            0 & 0 & J_{r} &  \epsilon_r' + E_m^r & t_r & 0  & \sqrt{2} t'_r\\
            t'_l & t_l & t'_r & t_r & U_1+E_m^m & J_m & J_m \\
            \sqrt{2} t_l & 0 & \sqrt{2} t'_r & 0 & J_m &   U_1 & 0\\
            0 & \sqrt{2}t'_l &  0 & \sqrt{2} t'_r & J_m &   0  & U_1+2E_m^m\\
        \end{pmatrix}.
    \end{equation}
\end{widetext}
While this Hamiltonian is much more compact and easily readable than the one applied in the main text, we will use it exclusively for the upcoming derivations of the ground states and corrected energy splittings of the system. The reason for this is that the exchange terms appearing in the (0,2,0)-charge sector such as $J_m$ in Eq.~\eqref{eq: Singlet exchange terms} will diverge during the charge transitions, making it appropriate far from the avoided crossing but inadequate for describing the  transitions that are vital for the protocol. For small $E_m$ we find that the exchange terms are of similar magnitude as the energy spacings $E_m^i$. Consequentially, the eigenstates of $H_S$ are not given  solely by states carrying zero, one, or two valley excitations, respectively, but by a mixture of those. As all our processes are assumed to be adiabatic, the electron will always stay in the lowest-energy eigenstate at each point of the protocol. Therefore, to obtain the approximate eigenstates at the relevant stages, we perform three additional Schrieffer-Wolff transformations onto $H_S$ to obtain an effective Hamiltonian for each of the three charge configuration regimes,
\begin{equation}
    \begin{aligned}
        H_{S,m} &= \begin{pmatrix}
            \tilde{U}_1+\tilde{E}_m^m & \tilde{J}_m & \tilde{J}_m  \\
            \tilde{J}_m & \tilde{U}_1 & 0\\
            \tilde{J}_m & 0 & \tilde{U}_1+2 \tilde{E}_m^m \\
        \end{pmatrix},   \\
        H_{S,i} &= \begin{pmatrix}
            \tilde{\epsilon}_i & \tilde{J}_i  \\
            \tilde{J}_i & \tilde{\epsilon}_i+\tilde{E}_m^i \\ 
        \end{pmatrix}.    
    \end{aligned}
\end{equation}
Here we once again, for the purpose of simplification, shifted the overall energy of the systems as $\tilde{\epsilon}_i = \epsilon_i'+ (2 t_i^2 + t_i'^2)/(\epsilon_i-U_1)$ and introduced the following effective parameters,
\begin{equation}
    \begin{aligned}
        \tilde{E}_m^m &\approx E_m^m - \frac{t_l'^{2}-t_l^{2}}{\epsilon_l-U_1} - \frac{t_r'^{2}-t_r^{2}}{\epsilon_r-U_1}, \\
        \tilde{J}_m &\approx J_m'- \sqrt{2} \left( \frac{t_l t_l'}{\epsilon_l-U_1}+ \frac{t_r t_r'}{\epsilon_r-U_1} \right),\\
        \tilde{J}_i &\approx J_i' + \frac{t_i t_i'}{\epsilon_i-U_1},\\
        \tilde{E}_m^i &= E_m^i + \frac{ t_i'^2 - t_i'^2}{\epsilon_i-U_1},\\
        \tilde{U}_1 &\approx U_1 -2 \frac{t_l^2}{\epsilon_l-U_1}- 2 \frac{t_r^2}{\epsilon_r-U_1}.
    \end{aligned}
\end{equation}
These individual Hamiltonians are easily diagonalizable leading to the following diagonal matrices and their associated unnormalized bases,
\begin{equation}
    \begin{aligned}
        &H_{S,m}^d = \begin{pmatrix}
            0 & 0 & 0  \\   
            0 & -E_{m,\text{eff}} & 0\\
            0 & 0 &  E_{m,\text{eff}} 
        \end{pmatrix}  +\tilde{U}_1+\tilde{E}_m',\\
        &H_{S,m}^b = \begin{pmatrix}
            -\frac{\tilde{E}_m'}{J_m} & \frac{-\tilde{E}_m^m - E_{m,\text{eff}}}{J_m}& \frac{-\tilde{E}_m^m + E_{m,\text{eff}}}{J_m}  \\   
            -1 & \frac{\left( \tilde{E}_m^m + E_{m,\text{eff}} \right)^2}{2 J_m^2} & \frac{\left( \tilde{E}_m^m - E_{m,\text{eff}} \right)^2}{2 J_m^2}\\
            1 & 1 & 1 
        \end{pmatrix} ,  
    \end{aligned} 
\end{equation}
and 
\begin{equation}
    \begin{aligned}
        &H_{S,i}^d = \begin{pmatrix}
            0 & 0  \\   
            0 &  E_{m,\text{eff}}^i 
        \end{pmatrix} + \tilde{\epsilon}_i + \frac{1}{2}\left( \tilde{E}_m^i- E_{m,\text{eff}}^i \right), \\
        &H_{S,(1,1,0)}^b = \begin{pmatrix}
            \frac{-\tilde{E}_m^i - E_{m,\text{eff}}^i}{2 J_i}& \frac{-\tilde{E}_m^i + E_{m,\text{eff}}^i}{2 J_i}  \\   
            1 & 1
        \end{pmatrix},
    \end{aligned} 
\end{equation} 
where the approximate middle-dot valley splitting in each sector is given by,
\begin{equation}
    \begin{aligned}
E_{m,\text{eff}} &\approx \sqrt{\left(\tilde{E}_m^m\right)^2 + 2 \tilde{J}_m^2 } ,\\
        E_{m,\text{eff}}^i &\approx \sqrt{\left( \tilde{E}_m^i \right)^2 + 4 \left( \tilde{J}_i \right)^2 }. 
    \end{aligned}
\end{equation}
Normalizing the basis vectors, we can apply the associated basis transformations to our Hamiltonian. The pertubation approach is only valid in the regime where $|\epsilon_i-U_1| \gg 0$, whereby this description only holds in the edge cases of our detuning protocol and would again diverge during the transitions. For the purpose of finding the eigenstates, we therefore fix $\epsilon_l=\epsilon_\text{start}$=const. for the transformation concerning the (1,1,0) regime, $\epsilon_l=\epsilon_r=\epsilon_\text{end}$ in the (0,2,0) regime, and $\epsilon_r=\epsilon_\text{start}$ in the (0,1,1) regime. Thereby, our transformation diagonalizes the system only at these specific detunings and we assume that the eigenstates during the transitions can be approximately estimated as the adiabatic mixtures of the eigenstates in that limit. 

We can perform an analogous procedure for the triplets where we will find the same Hamiltonian with the (1,1,0) energies shifted by the ordinary exchange splittings from the (2,0,0) to (1,1,0) transitions and a switch in signs in the exchange couplings of these sectors. Making the same simplifications as above, we obtain,
\begin{widetext}
    \begin{equation}
    \label{eq:Full-Exchange-Hamiltonian_triplet}
        H_{T} = \begin{pmatrix}
            \epsilon_l'+D_l & J_{l,T}  & 0 & 0 & t_l' \\
            J_{l,T} & \epsilon_l'+E_m^l +D_l'& 0 & 0 & -t_l \\
            0 & 0 & \epsilon_r' +D_r& J_{r,T} & -t'_r\\
            0 & 0 & J_{r,T} & \epsilon_r' + E_m^r+D_r'& t_r \\
            t'_l & -t_l & -t'_r & t_r & U_1+E_m^m  
        \end{pmatrix},
    \end{equation}
    with the definitions,
    \begin{equation}
        J_{i,T} = J_i + \frac{t_i t_i'}{ U_1+ \epsilon_i - E_m} + \frac{t_i t_i'}{ U_1+ \epsilon_i} \approx -J_i,  \hspace{5mm} D_i =  \frac{2 t_i^{2}}{ U_1+\epsilon_i}, \hspace{5mm} D_i' =  \frac{2 t_i'^{2}}{ U_1+ \epsilon_i - E_m} .
    \end{equation}
\end{widetext}
Note that we have here implemented the same global energy shift by $J_6^d$ and redefinitions of the parameters as in the singlet case. Not only do we see that the triplets are shifted in energy compared to their singlet counterparts but that the energy difference between states having an excited middle dot valley state has changed. This nonuniform difference in diagonal elements will lead to triplets having a slightly different admixture of (1,1,0) or (0,1,1) states in their eigenstates and a more sophisticated $S$-$T_0$ splitting. Finally, we do not care about excited middle-dot valley contributions at any step of the protocol, as we are only interested in the mobile electron and the phase it collects in the process of hopping through the TQD. Performing again a Schrieffer-Wolff transformation into the individual charge sectors gives,
\begin{widetext}
    \begin{equation}
        H_{T,i} = \begin{pmatrix}
            0 & 
            \tilde{J}_{i,T} \\
            \tilde{J}_{i,T} & \tilde{E}_m^i+\tilde{D}_i \\ 
        \end{pmatrix}   + \overbrace{\tilde{\epsilon}_i +\frac{2 t_i^2}{\epsilon_i+U_1}-\frac{2 t_i^2}{\epsilon_i-U_1}}^{= \tilde{\epsilon}_{i,T}},
    \end{equation}
    \begin{equation}
            \tilde{J}_{i,T} =
            \tilde{J}_{i}+\frac{2 t_i t_i'}{ U_1+ \epsilon_i} - \frac{2  t_i t_i'}{ \epsilon_i- U_1},  \hspace{5mm} 
            \tilde{D}_i =2\frac{ t_i^{2}- t_i'^{2}}{ \epsilon_i - U_1} - 2\frac{ t_i^{2} - t_i'^{2}}{ \epsilon_i + U_1},
    \end{equation}
    \begin{equation}
        \begin{aligned}
             H_{T,i}^d &= \begin{pmatrix}
                0 & 0  \\   
                 0 & E_{m,\text{eff}}^{i,T} \\
            \end{pmatrix}  + 
            \tilde{\epsilon}_{i,T}
            + \frac{1}{2}\left( \tilde{E}_m^i+\tilde{D}_i - E_{m,\text{eff}}^{i,T} \right), \hspace{5mm}
            H_{T,i}^b = \begin{pmatrix}
                \frac{\left( -\tilde{E}_m^i-\tilde{D}_i - E_{m,\text{eff}}^{i,T} \right)}{2 \tilde{J}_{i,T}}& \frac{\left( -\tilde{E}_m^i-\tilde{D}_i + E_{m,\text{eff}}^{i,T} \right)}{2 \tilde{J}_{i,T}}  \\   
                1 & 1
            \end{pmatrix},
        \end{aligned} 
    \end{equation}
    \begin{equation}
        E_{m,\text{eff}}^{i,T} = \sqrt{\left(\tilde{E}_m^i+\tilde{D}_i \right)^2+4 \tilde{J}_{i,T}^2}.
    \end{equation}
\end{widetext}
Lastly, the single $(0,2,0)$ triplet carries the same energy as its singlet counterpart, up to corrections from coupling between the different $(0,2,0)$ states which are of fourth order and thus negligible. Consequently, in the adiabatic regime, the singlet-triplet splitting at the plateau is given by,
\begin{equation}
    \Delta_{S-T_0} = E_{m,\text{eff}} . 
\end{equation}
The full, cumbersome formula for this value can be written as follows,
\begin{equation}
    \label{eq:full E_wait}
                \begin{aligned}
                 E_{\text{wait}} = & \bigg[2 \big(- \frac{\sqrt{2} t_l t_l'}{- 2 E_l + 2 E_m + 2 U - 2 \epsilon_l}  \\
                  & + \frac{\sqrt{2} t_l t_l'}{2 E_m + 2 U - 2 \epsilon_l} -  \frac{\sqrt{2} t_l t_l'}{- 2 E_l + 2 U - 2 \epsilon_l}  \\
                  & + \frac{\sqrt{2} t_l t_l'}{2 U - 2 \epsilon_l} - \frac{\sqrt{2} t_r t_r'}{2 E_m - 2 E_r + 2 U - 2 \epsilon_r} \\
                  & - \frac{\sqrt{2} t_r t_r'}{- 2 E_r + 2 U - 2 \epsilon_r} + \frac{\sqrt{2} t_r t_r'}{2 E_m + 2 U - 2 \epsilon_r}  \\
                  & + \frac{\sqrt{2} t_r t_r'}{2 U - 2 \epsilon_r}\big)^{2} + \big(E_m + \frac{t_l^{2}}{- E_l + E_m + U - \epsilon_l}  \\
                  & - \frac{t_l^{2}}{U - \epsilon_l} + \frac{t_l'^{2}}{E_m + U - \epsilon_l} - \frac{t_l'^{2}}{- E_l + U - \epsilon_l} \\
                  & + \frac{t_r^{2}}{E_m - E_r + U - \epsilon_r} - \frac{t_r^{2}}{U - \epsilon_r} \\
                  & - \frac{t_r'^{2}}{- E_r + U - \epsilon_r} + \frac{t_r'^{2}}{E_m + U - \epsilon_r}\big)^{2} \bigg]^{\frac{1}{2}} . 
                \end{aligned} 
\end{equation}
\section{Quasistatic charge noise during leapfrogging}
\label{App: Charge noise during leapfrogging}
Here we consider the effects of quasistatic charge noise in the gate voltages onto the leapfrogging protocol. To this end, we describe the tunneling parameters $(t_c)_{l,r}$ and detunings $\epsilon_{l,r}$ as being afflicted by some unknown deviations $\delta (t_c)_{l,r}$ and $\delta \epsilon_{l,r}$. 
This noise will lead to some finite dephasing effects captured by 
\begin{equation}
    \langle e^{i\delta \varphi} \rangle = e^{-\frac{1}{2}\langle \delta \varphi^2 \rangle},
    \label{eq: dephasing Q}
\end{equation}
in the case of zero-mean Gaussian noise. We assume the noise of different gates to be mostly uncorrelated, whereby $\langle \delta \epsilon \delta t_c \rangle = 0$ and the variance of the noise induced phase shift in the case of quasistatic noise simplifies to 
\begin{equation}
    \langle \delta \varphi^2 \rangle = \langle \delta \varphi_\epsilon^2 \rangle + \langle \delta \varphi_{t_c}^2 \rangle,
\end{equation}
where $\delta \varphi_\epsilon$ and $\delta \varphi_{t_c}$ are the phase contributions from the detuning and the tunnel coupling, respectively. Assuming fully adiabatic transitions, the qubits, being initialized in a superposition of the singlet and triplet ground states $\ket{LMS}$ and $\ket{LMT_0}$, will stay in the respective instantaneous ground state during the time evolution, allowing us to abbreviate the relevant states as $\ket{S}$ and $\ket{T_0}$.
Ignoring contributions of higher excited states, the ground-state energies can be modeled as those of regular LZ crossings; therefore,
\begin{equation}
        E_{S,i} = \frac{1}{2} \bigg( \epsilon_i - \frac{E_m}{2} - \overbrace{\sqrt{(\epsilon_i+\frac{E_m}{2})^2 + ( 2\sqrt{2} t_i)^2}}^{= \Omega_{S,i}} \bigg) 
\end{equation} 
\begin{equation}
    E_{T_0,i} = \frac{1}{2} \bigg( \epsilon_i + \frac{E_m}{2} - \underbrace{\sqrt{(\epsilon_i-\frac{E_m}{2})^2 + ( 2t_i')^2}}_{= \Omega_{T_0,i}} \bigg),
\end{equation}
where the definitions of the detunings are shifted to have $\epsilon_i = 0$ be centered between the singlet and triplet avoided crossings. In the following considerations, we will drop the index $i$ which denotes the left or right dot, as the noise dynamics of both transitions are identical and only vary in the choice of parameters. In addition, we define the effective singlet coupling as $t_e = \sqrt{2}t_i$ and $\alpha_i = \frac{t_i'}{\sqrt{2}t_i}$ as the renormalized ratio between inter- and intra-valley coupling. Consequently, the noisy energy difference is given by 
\begin{equation}
    \begin{aligned}
        \Delta E &= \frac{E_m}{2}-\frac{1}{2} \sqrt{(\epsilon+\delta \epsilon-\frac{E_m}{2})^2 + ( 2\alpha (t_e+\delta t_e))^2}\\ 
        &+\frac{1}{2} \sqrt{(\epsilon+\delta \epsilon+\frac{E_m}{2})^2 + (2(t_e+\delta t_e))^2},
    \end{aligned}
\end{equation}
which we can expand in terms of the small parameters $\delta \epsilon / \Omega_{S/T_0}$ and $\delta t_e / \Omega_{S/T_0}$, with the result,
\begin{widetext}
    \begin{equation}
        \varphi = \int_0^{t_f} \Delta E dt \approx \underbrace{\int_0^{t_f} \Delta E_0 dt}_{= \varphi_0} - \underbrace{\int_0^{t_f} \frac{ \delta \epsilon}{2} \left( \frac{\epsilon - \frac{E_m}{2}}{\Omega_{T_0}}-\frac{\epsilon + \frac{E_m}{2}}{\Omega_{S}} \right) dt}_{= \delta \varphi_{\epsilon}} - \underbrace{\int_0^{t_f} 2\delta t_e \left( \frac{\alpha^2 t_e}{\Omega_{T_0}}-\frac{t_e}{\Omega_{S}} \right) dt}_{= \delta \varphi_{t_c}}.
    \label{eq: full phase fluctuation}
    \end{equation}
\end{widetext}
Here, $\varphi_0$ corresponds to the deterministic noiseless dynamical phase resulting from the detuning process. As we are investigating quasistatic noise, $\delta\epsilon$ and $\delta t_e$ are time independent and reduce to a prefactor of their respective integrals. 
\subsection{Linear sweep}
Evaluating these integrals for a linear sweep $\epsilon(t) = v t$, we obtain,
\begin{widetext}
    \begin{gather}
        \delta \varphi_{\epsilon} = \frac{\delta \epsilon}{2v} \left[ \sqrt{(\epsilon-\frac{E_m}{2})^2+( \alpha t_e)^2}-\sqrt{(\epsilon+\frac{E_m}{2})^2+( t_e)^2} \right]^{\epsilon_{\text{end}}}_{\epsilon_{\text{start}}}, \\ 
        \delta \varphi_{t_c} = \delta t_e \frac{2t_e}{v} \left[\ln \left(\frac{1}{2t_e}\left(\Omega_S-\epsilon-\frac{E_m}{2}\right)\right)-\alpha^2\ln\left(\frac{1}{2\alpha t_e}\left(\Omega_T-\epsilon+\frac{E_m}{2} \right)\right)\right]^{\epsilon_{\text{end}}}_{\epsilon_{\text{start}}}.
    \end{gather}
\end{widetext}
For most realistic parameter ranges, the error caused by noise in the tunneling coefficient will be reasonably small with values $\langle\delta \varphi_{t_c}^2\rangle \leq 0.003 $, such that it does not endanger the feasibility of the operation to a large degree. The error scales linearly with the tunneling energy and strongly depends on the singlet-triplet coupling ratio $\alpha$. Particularly for tunnel couplings fulfilling $\alpha \approx 1$ the error will even vanish when performing a symmetric sweep, i.e., $\epsilon_{\text{end}} = \epsilon_{\text{start}}$. Therefore, it is advisable to perform leapfrogging at sites with a stronger intervalley than intravalley coupling. To determine that value, one requires knowledge about the local differences in valley phases. There are known ways to obtain the required parameters given a TQD system \cite{Russ2020-fh} and the ability of mapping out the valley phase across a two-dimensional semiconductor interface was recently demonstrated experimentally \cite{valleyphasemapping}. Thus, this requirement for $\alpha$ is justified. Such mapping procedures are not essential for the protocol though, as statistically, most TQDs will display sufficient values for $\alpha$ to guarantee feasibly low error rates.
%

In contrast, the noise coming from the fluctuation in the detunings, if left untreated, can be detrimental. For large sweeps $\epsilon_{\text{end}} \gg t_c$ the phase fluctuation coming from the detuning noise can be approximated as
\begin{equation}
    \delta \varphi_{\epsilon} = \frac{\delta \epsilon }{v}E_m,
\label{eq: detuning phase}
\end{equation}
and, consequentially, the dephasing linearly scales with both the noise intensity as well as the middle-dot valley splitting $E_m$. This makes sense, as the quasistatic noise shifts the time of the charge transition. While the singlet-triplet splitting in the (1,1,0) configuration is very small, as it is determined by just the residual exchange $\Delta_{S-T}(\epsilon_{\text{start}}) \approx \frac{4U_1t_l^2}{U_1^2-\epsilon_{\text{start}}^2} \sim 10^{-1} \mu \mathrm{eV}$, it is equal to the valley splitting in the (0,2,0) configuration $\Delta_{S-T}(\epsilon_{\text{end}}) \approx E_m$. As a result, if the electron transitions earlier or later than anticipated, then it is going to collect an additional phase equal to the energy difference $E_m$ multiplied by the uncertainty in the transition time $\delta t = \frac{\delta \epsilon}{v}$. 

As a consequence of equation~\eqref{eq: detuning phase}, we see that even for extremely low valley splittings $E_m$, there will be significant dephasing caused by quasistatic noise.
\subsection{Two-speed-sweep}
A method to reduce the impact of detuning noise onto dephasing is by choosing a velocity profile such that the integrals in equation \eqref{eq: full phase fluctuation} vanish. As the noise in the tunneling parameter is significantly weaker than the detuning noise and essentially does not scale with the valley splitting, we will focus on minimizing the integral corresponding to $\varphi_\epsilon$.
The simplest ansatz is to have a single velocity switch happen during the pulse, such that the dephasing is described by, 
\begin{equation}
    \begin{aligned}
        &\frac{2 \delta \varphi_{\epsilon}}{ \delta \epsilon}  =  \int_0^{t_f} \overbrace{\left( \frac{\epsilon - \frac{E_m}{2}}{\Omega_{T_0}}-\frac{\epsilon + \frac{E_m}{2}}{\Omega_{S}} \right)}^{= f_{S-T}(t)} dt \\
        &= \frac{1}{v_1}\int_{\epsilon_{\text{start}}}^{\epsilon_{\text{switch}}} f_{S-T}(\epsilon) d\epsilon + \frac{1}{v_2}\int_{\epsilon_{\text{switch}}}^{\epsilon_{\text{end}}} f_{S-T}(\epsilon) d\epsilon .
    \end{aligned}
\end{equation}
Examining the integrand $f_{S-T}$, one finds that it has both a positive and a negative sign in different domains. The value of epsilon at which the sign flips  and $f_{S-T}(\xi) = 0$ can be determined algebraically as,
\begin{equation}
    \xi = \frac{E_m}{2} \left( \gamma + \sqrt{\gamma^2 - 1} \right), \hspace{3mm} \gamma = \frac{\alpha^2+1}{1-\alpha^2}.
    \label{eq: epsilon switch}
\end{equation}
Consequentially, we can switch the velocities at this detuning $\xi$ and match $v_1$ and $v_2$ in such a way that the positive and negative regions are stretched so that the integrals of both cancel out. Therefore,
\begin{equation}
    \frac{1}{v_1} \left[ \Omega_T-\Omega_S \right]_{\epsilon_{\text{start}}}^\xi = \frac{1}{v_2} \left[ \Omega_T-\Omega_S \right]^{\epsilon_{\text{end}}}_\xi ,
\end{equation}
delivers the required ratio of the two velocities,
\begin{equation}
    r_v = \frac{v_2}{v_1} = \frac{\left[ \Omega_T-\Omega_S \right]^{\epsilon_{\text{end}}}_\xi}{\left[ \Omega_T-\Omega_S \right]_{\epsilon_{\text{start}}}^\xi}.
\end{equation} 

This provides a path towards significantly reducing the dephasing due to quasistatic noise in the detuning. The size of the negative region scales reciprocally with the valley splitting $E_m$, and thus, for larger $E_m$, one would need to choose ever slower velocities $v_2$ to cancel out the noise contributions, which is only practically feasible for low valley splittings. Additionally, the approximation of the energies, being those of single LZ crossings, becomes less accurate with higher valley splitting, whereby a more sophisticated method for identifying the optimal velocity profile would be required. As a result, this procedure enables good error suppression for operations, but does not extrapolate to arbitrarily high valley splittings. 

When doing a speed switch during the transition, one also needs to consider the effects this might have on other error sources. While a speed switch will not strongly impact the transfer fidelity, it will lead to an increase of the tunnel dephasing $\delta\varphi_{t_c}$ by a factor of roughly $\frac{1}{2}(1+ \frac{v_1}{v_2})$. Still, the dephasing induced by noise in the tunneling energies will be manageable, and multiplying the associated fidelity with the one obtained from the simulation for both parameter sets one stays above the error correction threshold.

During the waiting period of the leapfrogging protocol, the detuning will be held constant and is no longer dependent on time, but only on the energy difference $E_{\text{wait}}$ of the two relevant states at the waiting point, such that $t_{\text{wait}} = \frac{\pi}{E_{\text{wait}}}$, whereby,
\begin{equation}
    \begin{aligned}
        \delta \varphi_{\text{wait}} &= \int_0^{t_{\text{wait}}} \frac{ \delta \epsilon}{2} \left( \frac{\epsilon_{\text{end}} - \frac{E_m}{2}}{\Omega_{T_0,\text{end}}}-\frac{\epsilon_{\text{end}} + \frac{E_m}{2}}{\Omega_{S,\text{end}}} \right) dt \\
        &=    \frac{ \delta \epsilon}{2} \left( \frac{\epsilon_{\text{end}} - \frac{E_m}{2}}{\Omega_{T_0,\text{end}}}-\frac{\epsilon_{\text{end}} + \frac{E_m}{2}}{\Omega_{S,\text{end}}} \right) \frac{\pi}{E_{\text{wait}}}.
    \end{aligned}
\end{equation}
The full phase fluctuation collected during the leapfrogging protocol due to quasistatic noise is thus given by
    \begin{equation}
        \delta \varphi = \delta \varphi_{\text{wait}} + \delta \varphi^l + \delta \varphi^r  ,
        \label{eq: full phase fluctuation}
    \end{equation}
where $\delta \varphi^i = \delta \varphi_{\epsilon}^i + \delta \varphi_{t_c}^i$ with $i = l,r$ denote the fluctuations arising from the transition between the left and the middle, as well as the middle and the right dot, respectively.

\section{Obtaining the Gate Fidelity from the Dephasing}
Focusing here only on the dephasing, we consider the action of the Hamiltonian in the spin basis $\left\{ \ket{T_-}, \ket{T_0}, \ket{T_+}, \ket{S}\right\}$. For the calculation of the gate fidelity, we will assume the wait time to be tuned so that the leapfrogging implements an identity gate on the spin degree of freedom whereby the ideal unitary $U$ and its noisy implementaton $\tilde{U}$ become
\begin{equation}
    U = \begin{pmatrix}
            1 & 0 & 0 & 0 \\   
            0 & 1 & 0 & 0 \\
            0 & 0 & 1 & 0 \\
            0 & 0 & 0 & 1 \\
        \end{pmatrix} \\
        , \hspace{5mm}\tilde{U} = \begin{pmatrix}
            1 & 0 & 0 & 0 \\   
            0 & 1 & 0 & 0 \\
            0 & 0 & 1 & 0 \\
            0 & 0 & 0 & e^{i\delta\varphi} \\
        \end{pmatrix}.  \\
\end{equation}
For any $d$-dimensional qubit system the gate fidelity of a unitary operation is given by \cite{Fidelity_Dephasing}
\begin{equation}
\label{eq: fidelity}
    \mathcal{F_{\delta\varphi}} = \frac{d+\left|\mathrm{tr}\left(U^{\dagger}\tilde{U}\right)\right|^2}{d(d+1)}.
\end{equation}
Explicitly calculating the second term in the numerator therefore leads the following formula:
\begin{equation}
    \mathcal{F_{\delta\varphi}} = \frac{d+(d-1)^2+1+2(d-1)\cos{(\delta\varphi)}}{d(d+1)}.
\end{equation}
To obtain the average gate fidelity we still need to average over all noise implementations. With the help of equation \eqref{eq: dephasing Q} the average of the cosine term can be computed to be 
\begin{equation*}
    \langle \cos(\delta\varphi)\rangle = \frac{1}{2}\left( \langle e^{i\delta\varphi}\rangle + \langle e^{-i\delta\varphi}\rangle \right) = e^{-\frac{\langle \delta\varphi^2\rangle}{2}} 
\end{equation*}
whereby the average gate fidelity of the dephasing in $d=4$ case is given by
\begin{equation}
    \mathcal{F}_{\varphi} = \frac{7}{10}+\frac{3}{10}e^{-\frac{\langle \delta\varphi^2\rangle}{2}}.
    \label{eq: dephasing fid}
\end{equation}
The calculations for all other SWAP$^\gamma$ gates follows analogously and lead to the same result.
\section{Operation Intrinsic Errors}
There are also errors that are intrinsic to the leapfrogging operation and not related to noise. They arise from  leakage due to nonadiabaticity or incomplete transitions and depend strongly on the choice of operation parameters $\epsilon_{\text{end}}, \epsilon_{\text{start}}, v$. They limit the choice of those parameters to certain regions that deliver feasible error rates.

Since we assume the input quantum state to be pure, we can quantify the  of each such mechanism through the state-fidelity $Q_i$ defined as
\begin{equation}
    Q_i = 1-| \bra{\psi_f} \mathcal{E}_i (\ket{\psi}\bra{\psi}) \ket{\psi_f} |
\end{equation}
for any input state $\ket{\psi}$ and targeted final state $\ket{\psi_f}$, where $\mathcal{E}_i$ is the quantum channel associated with that error. As these errors arise from leakage rather than dephasing, we can consider their action onto the basis states independently and then choose the worst possible outcome as our estimate of infidelity from that operation.\\
From Appendix \ref{App: Charge noise during leapfrogging} we see that the error rate highly depends on the choice of $\epsilon_{\text{start}}$ and $\epsilon_{\text{end}}$. Setting $\epsilon_{\text{start}} = -\epsilon_{\text{end}}$ for a roughly symmetric sweep, we want to choose $\epsilon_{\text{end}}$ in such a way as to minimize the impact of noise during the whole protocol. Putting away at first the fact that a certain sweep range is needed to transition fully into the desired (0,2,0)-state, it is clear that the transition noises $\delta \varphi_{\epsilon,s}^l,\delta \varphi_{\epsilon,s}^r $ from going in and out of the middle dot are minimized by choosing $\epsilon_{\text{end}}$ to be as small as possible. On the other hand, choosing a large $\epsilon_{\text{end}}$ will minimize the impact of noise during the waiting period, as the energies sit on a plateau and are therefore insusceptible to change in $\epsilon$. The optimal value for $\epsilon_{\text{end}}$ is therefore found by minimizing \eqref{eq: full phase fluctuation}.
\begin{figure}
    \centering
    \includegraphics[width=\linewidth]{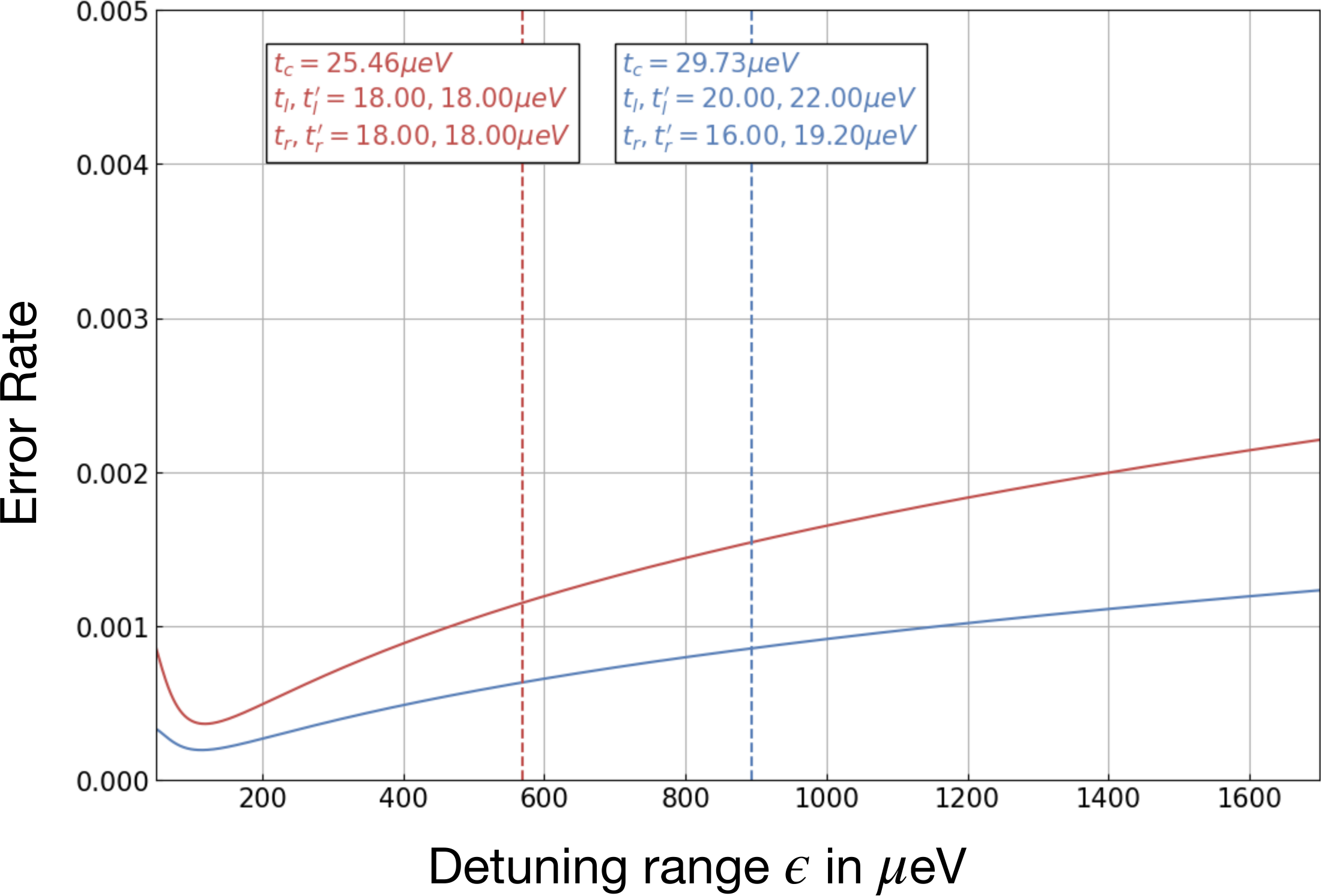} 
    \caption{Estimated infidelity resulting from dephasing and nonadiabaticity, depending on the size of the detuning sweep $\epsilon_{\text{end}}$ for the two parameter sets 1 (blue) and 2 (red). The vertical lines mark the detuning where the expectation value to find the mobile electron in the middle dot (i.e., for the whole configuration to be in the (0,2,0)-state) hits $99.9\%$.}\label{fig:full_transition}
\end{figure}
We find numerically that for typical tunnel barriers and detuning speeds that also guarantee sufficient adiabaticity, the optimal $\epsilon_{\text{end}}$ is very close to zero, which can be read from Figure~\ref{fig:full_transition}. Such low values are impractical, as there is a certain minimum value for $\epsilon_{\text{end}}$ such that the electron has transitioned sufficiently far enough into the (0,2,0)-state so that it can then be loaded into the right dot afterwards. 
Assuming full adiabaticity, the adiabatic ground state of a simple Landau Zener transition is given by,
\begin{equation}
    \ket{\tilde{S}}_{l} = \cos(\theta_{S,l}) \ket{S_0(1,1,0)} + \sin(\theta_{S,l}) \ket{S_0(0,2,0)},
\end{equation}
with $\theta_{S,l}$ being the adiabatic mixing angle for singlets having their mobile electron transitioning from the left to the middle dot.
While this angle can be calculated analytically for the simple LZ case, in our more complex situation, it suffices to see that the coefficients scale as ,
\begin{equation}
    (Q_{\text{trans}})_{S}^l = \cos^2(\theta_{S,l}) \approx  \frac{2t_l^2} {\epsilon^2} \overset{!}{<} 0.001,
\label{eq: condition for transmission error}
\end{equation}
for large detunings $\epsilon \gg 2\sqrt{2}t_l$. Consequentially, we expect to find a transmission error fulfilling the bound of $(Q_{\text{trans}})_{S}^l < 0.001$ at detunings around $500 \,\mu \text{eV} <\epsilon_{end}<1000\,\mu \text{eV}$ for most realistic parameter sets. We chose this specific bound as it is demonstrative and detuning much further than that will only lead to diminishing returns while other noise sources might increase at a quicker rate. As the singlets are usually coupled more strongly, they require a higher detuning to satisfy condition \eqref{eq: condition for transmission error} and will therefore display a higher transition error rate than their triplet counterparts. Hence, we pessimistically define $(Q_{\text{trans}})_{S}^i = Q_{\text{trans}}^i$ to stay conservative in our error estimation and the full transmission error is then calculated as the sum of the error of both transitions.
In Fig.~\ref{fig:full_transition} we show the detuning noise for two different sets of realistic parameters, as well as the minimum  $\epsilon_{\text{end}}$ to fulfill the necessary condition formulated in Eq.~\eqref{eq: condition for transmission error}.

The final error we consider is qubit loss due to a small non-adiabaticity of the transition, which is given by the Landau-Zener formula~\cite{non-adiabatic-landau-zener},
\begin{equation}
    (Q_{\text{LZ}})_{T_0} \approx  \exp \left( -2 \pi \frac{\alpha_l^2 t_l^2}{2 v} \right) +\exp \left( -2 \pi \frac{\alpha_r^2 t_r^2}{2 v} \right).
\end{equation} 
Here we use the LZ-error of the triplet as it will usually be the larger of the two. While the error by incomplete transition determines the range of the detuning sweeps, the error from nonadiabaticity limits the allowed sweep velocity. As can be seen from Eq.~\eqref{eq: detuning phase}, fast sweeps are generally perferrable. Thus, the choice of the detuning velocity is a compromise, minimizing both the nonadiabaticity and transition errors as much as possible \footnote{Additionally, a faster detuning speed will also reduce the gate time, which is not only preferable from an operational point of view, but also minimizes errors from unaccounted for dephasing and spin-flip processes.}.

\section{Full Fideltiy estimation}
Finally, assuming that each of the errors considered is sufficiently small and uncorrelated, we can formulate the total error of the leapfrogging sequence as the product of each of the individual error sources infidelities. Therefore,
\begin{equation}
    \mathcal{F} = \mathcal{F}_\varphi (1-Q_{\text{LZ}}) (1-Q_{\text{trans}}),
    \label{eq: full error rate}
\end{equation}
where $\mathcal{F}_\phi$ is the fidelity accounting for dephasing errors arising from the charge noise in the detuning and tunneling parameters $\delta \epsilon$ and $\delta t_c$, calculated as in Eq.~\eqref{eq: dephasing fid}. 
\begin{figure}
    \centering
    \includegraphics[width=\linewidth]{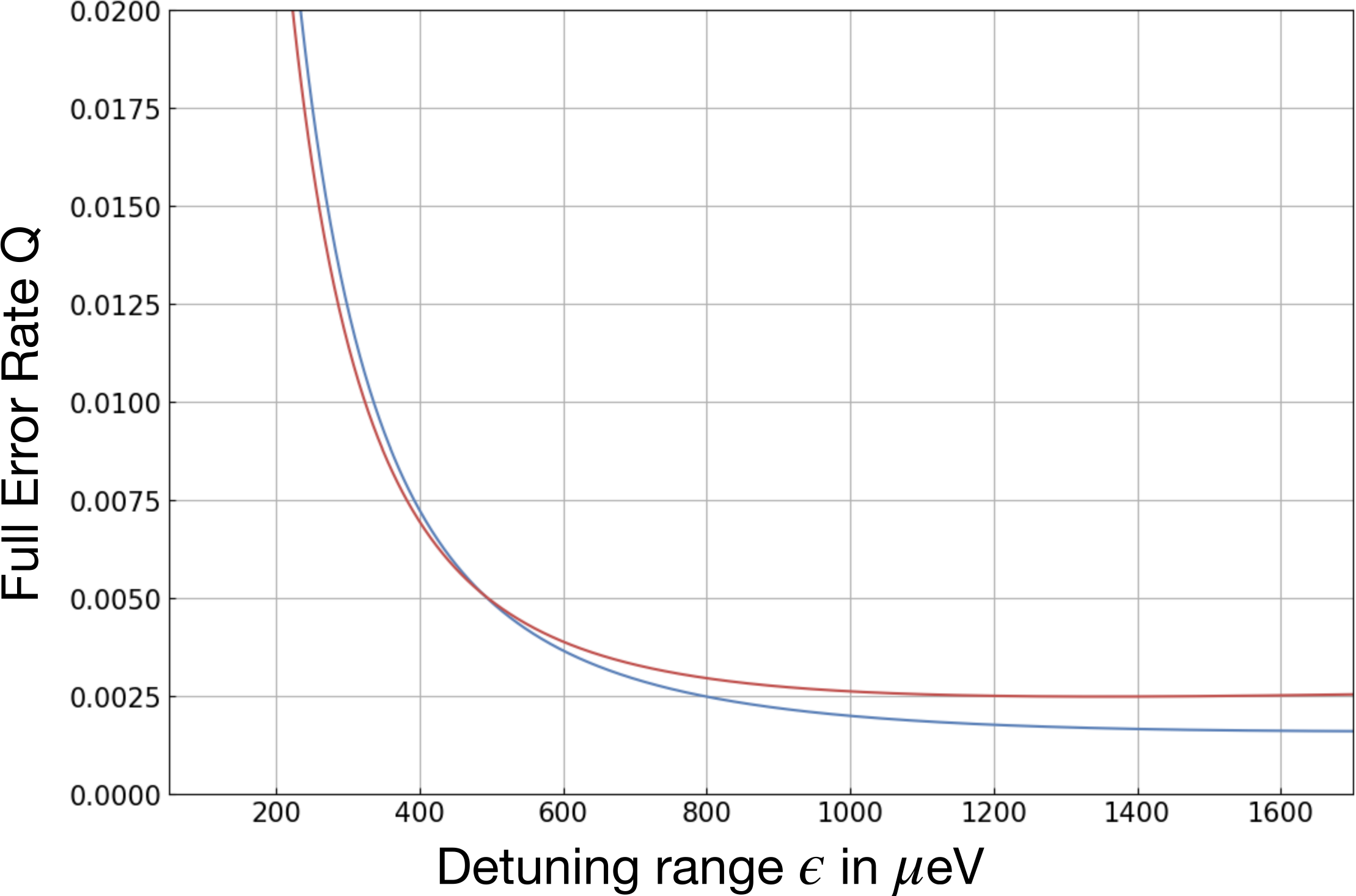} 
    \caption{ Estimated full error rate resulting from dephasing, nonadiabaticity and incomplete transmission depending on the size of the detuning sweep $\epsilon_{\text{end}}$ for the two parameter sets 1 (blue) and 2 (red). The minimum of these curves mark the range of the detuning sweep that optimises the protocol.}
    \label{fig:full_error}
\end{figure}
Assuming a noise spectral density of $S(1 \text{Hz}) \approx 1 \,\mu \text{eV}^2/{\text{Hz}}$ the noise level in the detunings can be estimated as $\langle \delta \epsilon^2 \rangle = 4.4 \,\mu \text{eV}$, which agrees with what is measured experimentally \cite{Connors2022-ik}. As there is no need for any controllability in the tunneling gates, their corresponding dc signal can be heavily filtered \cite{Filtered1, Filtered2} and one can expect their noise level to be significantly lower than usual, and thus we expect a tunneling noise amplitude of $\langle \delta 2t_c\rangle = 0.05\, \mu \text{eV}$. Using all of these considerations, we can estimate the errors for both parameter sets listed in the main text as $1-\mathcal{F}_1 \approx 1.70 \times 10^{-3}$ and $1-\mathcal{F}_2 \approx 2.62 \times 10^{-3}$, exceeding the surface code threshold of $\mathcal{F} \geq 0.99$ and leaving enough error budget for otherwise unaccounted for error mechanisms. The full estimated error rate in dependence on $\epsilon_{\text{end}}$ is plotted in Fig.~\ref{fig:full_error}, where the minimum corresponds to the optimal detuning range to minimize the product of incomplete transition and dephasing errors. From this one can see that the fidelity plateaus around $\epsilon_{\mathrm{end,1}} = 1300 \,\mu \mathrm{eV}$ for the first and $\epsilon_{\mathrm{end,2}} = 1000 \,\mu \mathrm{eV}$ for the second parameter set.  We expect the realistic error from the discussed mechanisms to be of a similar scale to what is estimated here, up to some differences due to the slightly more complex energy transitions and the resulting mistiming of the speed switch.

\section{Effect of spin-orbit coupling and B-field gradients}
\label{App:SO-Grad}
Here, we discuss how finite $B$-field gradients and intrinsic spin-orbit coupling affect the leapfrogging protocol. This also includes dot-to-dot $g$-factor differences, as their effect is identical to that of magnetic field gradients. 

\subsection{Magnetic field gradients}
Differences in $B$-fields in between the dots change the protocol in two distinct ways depending on the direction of the gradient 
(longitudinal $\Delta B_\parallel$ or transversal $\Delta B_\perp$). Transversal magnetic field gradients $\Delta B_\perp$ couple the spin singlets $\ket{S}$ with the polarized spin triplets $\ket{T_\pm}$ in the singly occupied charge regimes and during transitions with coupling strength,
\begin{equation*}
    \bra{T_\pm}H\ket{S} = \mp\frac{g\mu_B \Delta B_\perp}{2\sqrt{2}},
\end{equation*}
which for conservative estimates of $\Delta B_\perp < 0.3$ mT \cite{architecture1} amounts to a coupling on the order of $\sim 10^{-2} \,\mu$eV. Although not leading to any meaningful valley-orbit coupling, their main effects are a slight mixing of the spins in the (1,1) eigenstates, as well as the introduction of possible avoided crossings during the charge transition. Depending on the the valley splitting and Zeeman energy of the middle dot, there can exist possible regimes where avoided crossings between the singlet $\ket{S}$ and negatively polarized triplet $\ket{T_-}$ occur. For the proposed detuning velocities of $v \approx 100 \,\mu \text{eV/ns}$, the Landau-Zener formula predicts a leakage probability of roughly,
\begin{equation}
    P_{S\rightarrow T_-} \approx 1-\exp\left({-2\pi \frac{\left(g\mu_B \Delta B_\perp\right)^2}{16v}}\right) \approx 7.2 \times 10^{-6},
\end{equation}
which is negligible for our purposes.

The resulting contribution of polarized triplets in the ``mostly-singlet''-states is given to first order by,
\begin{equation}
    \ket{\tilde{S}} \approx \ket{S} + \sum_\pm \frac{\bra{T_\pm}H\ket{S}}{E_S-E_{T_\pm}}\ket{T_\pm},
\end{equation}
where the stated energy difference is calculated as $E_S-E_{T_\pm} = J \pm E_Z$. Thus, the contribution of each triplet is,
\begin{equation*}
    c_\pm = \frac{\pm g \mu_B \Delta B_\perp}{2\sqrt{2}(J \pm E_Z)},
\end{equation*}
which for $g \approx 2$ and $J\pm E_Z \approx 10 \, \mu $eV will be on the order of $c_\pm \sim 10^{-3}$,
and thus largely negligible. Even if it was of significant proportion, since the detuning and shuttling is done adiabatically compared to the spin-preserving anticrossing, while skipping over the non spin-preserving ones (i.e., the $S-T_-$-anticrossing), one preserves the superposition representing the initial spin configuration up to a SWAP$^\gamma$-gate and additional deterministic phases on the ``mostly polarized triplet''-states, which can again be corrected digitally. Thus, one would afterwards adiabatically transition back into the desired spin states, regardless of the mixture of singlets and triplets in the eigenstates of the Hamiltonian. Due to the $T_-$-triplet having a stronger mixing with the singlet than the $T_+$-triplet, there will be a small difference in energy deviation and hence differing phase accumulation of the polarized triplets and the $T_0$-triplet during the early detuning stage, which cannot be digitally corrected. Introducing $t_\perp = \frac{g\mu_B \Delta B_\perp}{2\sqrt{2}}$ as the coupling, the energy difference is calculated as,
\begin{widetext}
    \begin{equation}
        \Delta E_{T_+,T_-} =
        E_Z +\frac{1}{2}\left( \sqrt{(J+E_Z)^2+4t_\perp^2}-\sqrt{(J-E_Z)^2+4t_\perp^2}\right) 
        \approx 2E_Z+t_\perp^2\left( \frac{1}{J+E_Z}-\frac{1}{J-E_Z}\right),
    \end{equation} 
\end{widetext}
where we used that for the expected case of small gradients $4t_\perp^2 \ll J^2, E_Z^2$ one finds that $\sqrt{(J\pm E_Z)^2+4t_\perp^2} \approx (J\pm E_Z)+2\frac{t_\perp^2}{J \pm E_Z}$. As a result, one obtains a deviation from the expected energy difference of $2E_Z$ of,
\begin{equation}
    \delta E_{T_+,T_-} \approx \frac{E_Zt_\perp^2}{J^2-E_Z^2} \approx -\frac{t_\perp^2}{E_Z},
\end{equation}
which is on the scale of $\sim 10^{-5} \, \mu eV$ for the assumed values and should have no measurable impact on the protocol. Starting at a gradient of $\Delta B_\perp \approx 1 \, {\rm mT}$, one can expect leakage probabilities $P_{S\rightarrow T_-} \sim 10^{-4}$ as well as anisotropies of $\delta E_{T_+,T_-} \approx 10^{-2}\, \mu$eV to appear, where the introduced error would become a major source of infidelity and their handling would require further considerations.

Longitudinal gradients $\Delta B_\parallel$ and dot-to-dot differences in $g$-factors lead to a difference in the Zeeman energy $\Delta E_Z$ between two adjacent dots. As a result, there is an induced coupling between the spin singlet $\ket{S}$ and  the triplet $\ket{T_0}$ states. Compared to the previous case, these two states are closer in energy, only separated by the exchange $J$ which in the relevant singly occupied configurations is small. As a result, the new dressed eigenbasis in such a scenario is given by the states,
\begin{gather}
        \ket{\pm} = \cos (\frac{\theta}{2}) \ket{S}\pm \sin (\frac{\theta}{2}) \ket{T_0}, \\
        \tan (\theta) = \frac{ \Delta E_Z}{J},
\end{gather}
which results in $\left\{ \ket{S}, \ket{T_0}\right\}$ being the suitable eigenbasis for large $ J \gg \Delta E_Z$ and $\left\{ \ket{\uparrow\downarrow}, \ket{\downarrow \uparrow }\right\}$ for the opposite case. Assuming an exchange energy of $J \approx 0.1 \, \mu eV$ at the start of the protocol and a Zeeman energy difference of $\Delta E_Z \approx \, 5$MHz \cite{jones2026multiqubitentanglementunitcell}, one arrives at a singlet fraction in the $\ket{+}$-state of $1-c_S^+ \approx 5.19 \times 10^{-3}$. 
During the charge transition, $J$ will grow rapidly, whereby the basis will transition back to $\left\{ \ket{S}, \ket{T_0}\right\}$. By the same adiabaticity argument as for the transverse gradient, the makeup of the dressed states which the electrons occupy during the transition is not of particular relevance in itself, as the superposition of $\ket{\pm}$ is preserved. The coefficients of this superposition are strongly dependent on $\frac{J}{\Delta E_Z}$ and the additional unwanted operation in the pure spin basis is not as easily corrected as in the earlier case. Depending on this ratio, one can achieve a wide spectrum of two-qubit gates, ranging from SWAP$^\gamma$ operations to CPHASE$^\gamma$ gates, described in detail in \cite{NiPing25}. While it is difficult to asses the infidelity caused by this, one can, by first-order Trotterization, treat this as a SWAP gate and a weak  CZ gate happening sequentially without caring about the order for this rough estimation. As the regime where the magnetic field gradient is relevant can be estimated to be half of the duration of the transition, the additional phase on one of the qubits will be roughly $\Delta E_Z \times 30$~ns $\approx 0.15$ for the considered detuning speeds (see Fig.~\ref{fig:full_transition}). Applying the usual fidelity formula from Eq.~\eqref{eq: fidelity} yields a maximum infidelity of $1-\mathcal{F} \approx 9.66 \times 10^{-3}$ from an unaccounted Zeeman energy difference of the aforementioned scale. Although this is merely a rough estimation and higher order terms would likely cancel some of this error, this consideration implies that for high-fidelity operation one would certainly need to account for these effects. 
Usually, in the case of Si/SiGe Loss-DiVincenzo (LD) qubits, one desires to have a sufficiently large gradient to increase individual qubit addressability. Good data regarding unwanted stray magnetic field gradients is sparse and usually limited to exchange-only qubits, which have significantly different hardware and architectural requirements. With the rise of LD-architectures requiring such regions \cite{architecture1}, it can be expected that these stray fields can be reduced by a large factor through an optimization of the micromagnet design. This can in turn increase SWAP-fidelities in leapfrogging until it their fidelities are limited almost entirely by the dot-to-dot g-factor variation. Finally, it is important to note that, while this generally diminishes SWAP$^\gamma$-fidelities, the error is deterministic and can hence be accounted for in the gate operation. Since $J$ is tunable via the inter-dot detuning while $\Delta E_Z$ remains constant, the Hamiltonian offers universal control in the $S-T_0$-basis, corresponding to a high degree of control over the implemented two-qubit gate. By tuning the velocity profile to achieve only full $2 \pi$-rotations in the CZ-gate, one can, e.g., cancel the effect of the gradient completely, up to control imperfections. In particular, if one tunes the $B_z$-gradients sufficiently, one could also use leapfrogging to implement a CZ phase gate, thus switching to a fully different mode of operation. 

\subsection{Spin-orbit- and valley-orbit-coupling}
Spin-orbit-coupling (SOC) in Si/SiGe devices happens through Rashba- and Dresselhaus-like effects caused by inversion symmetry breaking at the interface \cite{BurkardReview2023}. It couples spin up and spin down states inside each dot by 
\begin{equation*}
    H_{\text{SO}} = \beta_D\left( \sigma_y P_y-\sigma_x P_x \right) + \alpha_R\left(\sigma_x P_y + \sigma_y P_x \right)
\end{equation*}
where $\alpha_R, \beta_D$ are the Rashba- and Dresselhaus-coefficients, $\sigma_{x,y}$ are the Pauli matrices along the crystallographic axes and $P_x$ and $P_y$ are the components of the generalized momentum vector. In Si/SiGe devices, SOC is typically quite small and leads only to a weak mixing of the eigenstates and thus implicit use of a dressed basis \cite{Cywinski-charge-transfer-decoherence}. This dressed basis still constitutes a Kramers doublet and, aside from a slight change in the renormalized Zeeman energy splitting, which we dealt with in the previous section, these states behave just as regular spin states. However, as the individual make-up of the dressed basis is site-dependent, this induces an overlap between the wavefunctions of different spin orientations in neighboring dots, thus enabling spin-flip tunnelling, which manifests itself as a finite matrix element $\bra{S(0,2)}H\ket{T_-(1,1)} =\Delta_{\text{SO}} \approx 10^{-1}\,\mu$eV  \cite{DanonPSBSOI2009,TanttuDzurak2019SOI,Raith_2012}. Additionally, SOI also induces another coupling mechanism, namely valley-orbit coupling, the interaction strength of which is usually stated to induce an anticrossing on a scale of around $\Delta_{\text{VO}} \approx 20 $~neV \cite{conveyor1}. It depends linearly on the valley splitting, whereby it is supposed to be much lower for the purpose of leapfrogging. Excluding the $\bra{S(1,1)}H\ket{T_-(1,1)}$ component, because it appears in second order to an already weak effect, summing and rounding up both of these provides a conservative upper bound for the size of the anticrossing of $\Delta_{S-T_-} \leq \Delta_{\text{SO}}+\Delta_{\text{VO}} \approx 150 $~neV. Plugging this into the Landau-Zener equation predicts a maximum SOI induced leakage probability of,
\begin{equation*}
    P_{S\rightarrow T_-} \leq 1-\exp\left({-2\pi \frac{\left(\Delta_{\text{SO}}+\Delta_{\text{VO}}\right)^2}{2v}}\right) \approx 6.9 \times 10^{-4}.
\end{equation*}
Although this is of relevant proportion, its magnitude does not threaten the feasibility of leapfrogging. Additionally, for the parameters proposed in this paper, there is no expected crossing of the $S-T_-$ energy levels, therefore removing this leakage channel as a whole. If such a crossing where to appear, one would expect threatening leakage probabilities of $P_{S\rightarrow T_-} \geq 1 \times 10^{-3}$ at a combined coupling of $\Delta_{\text{SO}}+\Delta_{\text{VO}} \gtrsim 250$~neV, requiring further consideration.

\bibliography{references.bib}

\end{document}